\pdfoutput=1
\documentclass[a4paper,11pt]{article}

\usepackage{jheppub} 

\usepackage[T1]{fontenc} 
\usepackage[section]{placeins}

\usepackage{graphicx}
\usepackage{mathtools}
\usepackage{amssymb}
\usepackage{amsfonts}
\usepackage[footnotesize]{caption}
\usepackage[font=scriptsize]{subcaption}
\usepackage{color}
\usepackage{braket}
\usepackage{hyperref}
\usepackage{url}
\usepackage{multirow}
\usepackage{relsize}
\usepackage{float}

\usepackage{tikz}
\usepackage{cancel}
\usepackage[compat=1.1.0]{tikz-feynman}
\usetikzlibrary{shapes,arrows}

\newcommand{\be}{\begin{equation}}
\newcommand{\ee}{\end{equation}}
\newcommand{\bea}{\begin{eqnarray}}
\newcommand{\eea}{\end{eqnarray}}

\newcommand{\deltaf}{\delta^{F}}
\newcommand{\deltat}{\delta^{T}}
\newcommand{\deltatt}{\delta^{TT}}
\newcommand{\deltaft}{\delta^{FT}}

\tikzstyle{decision} = [diamond, draw, fill=yellow!15, 
text width=6em, text badly centered, node distance=3cm, inner sep=0pt]
\tikzstyle{block} = [rectangle, draw, fill=blue!15, 
text width=12.5em, text centered, rounded corners, minimum height=2em]
\tikzstyle{block0} = [rectangle, draw, fill=red!30, 
text width=5em, text centered, rounded corners, minimum height=2em]
\tikzstyle{line} = [draw, -latex']
\tikzstyle{cloud1} = [draw, ellipse,fill=green!40, node distance=3cm,
minimum height=2em]
\tikzstyle{cloud2} = [draw, ellipse,fill=red!40, node distance=3cm,
minimum height=2em]
\tikzstyle{io} = [trapezium, trapezium left angle=70, trapezium right angle=110, minimum width=3cm, minimum height=1cm, text centered, draw=black, fill=blue!30]

\catcode`\@=11
\font\manfnt=manfnt
\def\Watchout{\@ifnextchar [{\W@tchout}{\W@tchout[1]}}
\def\W@tchout[#1]{{\manfnt\@tempcnta#1\relax%
  \@whilenum\@tempcnta>\z@\do{%
    \char"7F\hskip 0.3em\advance\@tempcnta\m@ne}}}
\let\foo\W@tchout
\def\dubious{\@ifnextchar[{\@dubious}{\@dubious[1]}}

\def\@dubious[#1]{%
  \color{red}\setbox\@tempboxa\hbox{\@W@tchout#1}
  \@tempdima\wd\@tempboxa
  \list{}{\leftmargin\@tempdima}\item[\hbox to 0pt{\hss\@W@tchout#1}]}
\def\@W@tchout#1{\W@tchout[#1]}
\catcode`\@=12

\preprint{LAPTH-023/18}
\title{\boldmath Non-minimal flavour violation in $A_4\times SU(5)$ SUSY GUTs
with smuon assisted dark matter}

\author[a]{Jordan Bernigaud\,}
\author[a]{\!\!\!,~ Bj\"orn Herrmann\,}
\author[b]{\!\!\!,~ Stephen F.\ King\,}
\author[b]{\!\!\!,~ Samuel J.\ Rowley\,}

\affiliation[a]{Univ.\ Grenoble Alpes, USMB, CNRS, LAPTh, 9 Chemin de Bellevue, F-74000 Annecy, France}
\affiliation[b]{School of Physics and Astronomy, University of Southampton, SO17 1BJ Southampton, United Kingdom}

\emailAdd{bernigaud@lapth.cnrs.fr}
\emailAdd{herrmann@lapth.cnrs.fr}
\emailAdd{king@soton.ac.uk}
\emailAdd{s.rowley@soton.ac.uk}

\abstract{
We study CP-conserving non-minimal flavour violation in $A_4 \times SU(5)$ inspired Supersymmetric Grand Unified Theories (GUTs), focussing on the regions of parameter space where dark matter is successfully accommodated due to a light right-handed smuon a few GeV heavier than the lightest neutralino. In this region of parameter space we find that some of the flavour-violating parameters are constrained by the requirement of the dark matter relic density, due to the delicate interplay between the smuon and neutralino masses. By scanning over GUT scale flavour violating parameters, constrained by low-energy quark and lepton flavour violating observables, we find a striking difference in the results in which individual parameters are varied to those where multiple parameters are varied simultaneously, where the latter relaxes the constraints on flavour violating parameters due to cancellations and/or correlations. Since charged lepton-flavour violation provides the strongest constraints within a GUT framework, due to relations between quark and lepton flavour violation, we examine in detail a prominent correlation between some of the flavour violating parameters at the GUT scale consistent with the stringent lepton flavour violating process $\mu \rightarrow e \gamma$. We also examine the relation between GUT scale and low scale flavour violating parameters, for both quarks and leptons, and show how the usual expectations may be violated due to the correlations when multiple parameters are varied simultaneously.
}


\begin{document}
\maketitle
\flushbottom


\section{Introduction}

Despite the absence of any direct experimental evidence, supersymmetric extensions continue to provide attractive solutions to the shortcomings of the Standard Model (SM). Phenomenologically, they provide viable dark matter candidates and can readily accommodate seesaw explanations for neutrino masses. From a more theoretical point of view, they cure the hierarchy problem related to the Higgs mass and lead to a more precise gauge-coupling unification as compared to the Standard Model (SM). This last point can be seen as a hint towards Grand Unified Theories (GUTs), where gauge couplings, certain soft-breaking parameters, and representations of SM fields are unified. 

The fact that the predicted superpartners have not been observed so far, may to some extent be moderated by the argument that current direct searches rely on specific assumptions, for example when all scalar masses are assumed to be degenerate at the Grand Unification scale. Moreover, as superpartner mass limits increase, the assumption of the Minimal Flavour Violation (MFV) paradigm postulating that the flavour structure of the theory is the same as in the SM, i.e.\ all flavour-violating interactions are related to the CKM- and PMNS-matrices only, may be relaxed without violating experimental limits. Relaxing this assumption and allowing for additional sources of flavour violation leads to a modification of the assumed decay patterns, for example in squark searches. As a further consequence, the obtained mass limits may be considerably weakened \cite{Blanke2015, LH2017, NMFVexp2018}. In addition, it appears that a considerable region of the parameter space of the TeV-scale Minimal Supersymmetric Standard Model (MSSM) can accomodate such Non-Minimal Flavour Violation (NMFV) in the squark sector with respect to current experimental and theoretical constraints \cite{Arana2014, Kowalska2014, NMFV2015}.

In recent years, the possibility of NMFV has received considerable attention in the context of studies of TeV scale physics \cite{Chung:2003fi}, mainly concerning collider signatures \cite{NMFV2005, NMFV2007a, NMFV2007b, NMFVReport2008, NMFV2009a, NMFV2009b, NMFV2010a, NMFV2010b, NMFV2011, NMFV2012, NMFV2014, NMFV2016, NMFV2017, NMFVRec2018}, but also dark-matter-related aspects \cite{Barger:2009gc, NMFVDM2011, NMFVDM2012, Blanke2014DM} and precision data \cite{Gabbiani:1988rb, Hagelin:1992tc, Gabbiani:1996hi, Dutta:2018fge}. Moreover, a considerable amount of work has been realized on flavour violation aspects within ultraviolet frameworks \cite{King:1998nv, Barenboim:2000ev, Baek:2001kh, Blazek:2002wq, Hayes:2005et, Calibbi:2006nq, Kim:2006ab, Cheung:2007pj, Ciuchini:2003rg, Ciuchini:2007ha, Antusch:2007re, Albright:2008ke, Masiero:2008cb, Albaid:2011vr, Moroi:2013vya, SU52014, SU52015}. In a Grand Unification framework, the flavour structure may be generated at the high scale, e.g., through flavour symmetries. Within a supersymmetric theory, imposing a certain symmetry implies a pattern of the soft-breaking terms which may accomodate flavour violating terms at the unification scale. Renormalization group running then generates the corresponding terms at the TeV scale, which enter the phenomenology at the TeV and electroweak scales. 

The link between NMFV terms at the TeV scale and the GUT scale is interesting from both the phenomenological and model-building point of view. First, although they may be numerically rather different, flavour violating interactions in the squark and slepton sectors are linked if NMFV is implemented in a unification framework at the high-scale. The same source of flavour violation may therefore be challenged by experimental data from both sectors. Second, gaining information on the flavour structure of a new physics framework at the TeV scale may give valuable hints towards Grand Unification and related flavour symmetries. The present paper is a first step in this direction.

In a recent paper by one of the authors \cite{A4SU5}, such a scenario was discussed in the framework of an $SU(5)$ GUT combined with an $A_4$ family symmetry\footnote{Note that the $A_4$ may be replaced by $S_4$ or $SO(3)$ or indeed any family symmetry which contains both triplet and singlet representations.}. The idea was that the three $\bf\overline{5}$ representations form a single triplet of the family symmetry with a unified soft mass $m_F$, while the three $\bf 10$ representations are singlets with independent soft masses $m_{T_1}$, $m_{T_2}$, $m_{T_3}$. Assuming MFV, it was shown that in order to account for the muon anomalous magnetic moment $(g-2)_{\mu}$, dark matter and LHC data, non-universal gaugino masses $M_i$ ($i=1,2,3$) at the high scale are required in the framework of the MSSM. 

We focussed on a region of parameter space that has not been studied in detail before characterised by low higgsino mass $\mu \approx -300$ GeV, as required by the $(g-2)_{\mu}$. The latter also required a right-handed smuon $\tilde\mu_R$ with a mass around 100 GeV, and a neutralino $\tilde{\chi}^0_1$ several GeV lighter which allows successful dark matter. The LHC will be able to fully test this scenario with the upgraded luminosity via muon-dominated tri- and di-lepton signatures resulting from higgsino dominated $\tilde{\chi}^\pm_1 \, \tilde{\chi}^0_2$ and $\tilde{\chi}^+_1 \, \tilde{\chi}^-_1$ production, as well as direct smuon production searches in the above region of parameter space.

The above study \cite{A4SU5} was clearly concerned with the implications of the flavoured GUT model for the superpartner spectrum consistent with $(g-2)_{\mu}$ and the Dark Matter relic density. However, for simplicity, it was assumed that there was no flavour violation at the GUT scale, whereas it is well known that such flavour violation is expected in these models \cite{Antusch:2013wn, Dimou:2015yng, Dimou:2015cmw}. The goal of the present study is to extend this work to the NMFV framework by introducing off-diagonal squark and slepton mass-squared terms in the Lagrangian at the GUT scale, motivated by the analyses in Refs.\  \cite{Antusch:2013wn, Dimou:2015yng, Dimou:2015cmw} which show that such flavour violation is generically expected. Here, we take a phenomenological (or model independent) approach, and simply introduce flavour violating terms at high energy to explore their effect on low energy observables. To this end we consider two MFV reference parameter points, one of which is inspired by the findings of Ref.\ \cite{A4SU5} and involves a very light smuon capable of accounting for $(g-2)_{\mu}$, and the other one with a heavier smuon, harder to discover at the LHC, but not able to account for $(g-2)_{\mu}$. In both cases, we then perturb around these points, switching on off-diagonal mass terms, consistently with $SU(5)$, arising from $A_4$ breaking effects. We find interesting correlations between the flavour violating parameters at the GUT scale consistent with the stringent lepton flavour violating processes $\mu \to e \gamma$.

The present paper is organised as follows: In Sec.\ \ref{Sec:Model}, we first introduce the $A_4 \times SU(5)$ supersymmetric model under consideration and present the implementation of NMFV therein. Sec.\ \ref{Sec:Setup} then presents the method and relevant tools for our analysis. Our results are presented and discussed in Sec.\ \ref{Sec:Results}. Finally, our conclusions are given in Sec.\ \ref{Sec:Conclusion}.


\section{Non-Minimal Flavour Violation in SUSY GUTs}
\label{Sec:Model}

In this Section, we present the model under consideration in this paper, namely the Minimal Supersymmetric Standard Model (MSSM) based on $SU(5)$ Grand Unification, including an $A_4$ flavour symmetry entailing NMFV terms already at the GUT scale. 

\subsection{SUSY-breaking in the MSSM}
Although the exact breaking mechanism is not completely understood, it is well known that Supersymmetry (SUSY) must be broken to some degree. The associated SUSY-breaking Lagrangian contains all terms which do not necessarily respect SUSY but hold to the tenets of gauge invariance and renormalisability. Considering the Minimal Supersymmetric Standard Model (MSSM), the SUSY-breaking Lagrangian reads
\begin{align}
	\begin{split}
		\mathcal{L}^{\rm MSSM}_{\rm soft} = 
		& - \frac{1}{2} \big( M_1\widetilde{B}\widetilde{B}+M_2\widetilde{W}\widetilde{W} 
							+ M_3\widetilde{g}\widetilde{g} + \rm{h.c.} \big) \\[1ex]
		& - M_Q^2\widetilde{Q}^{\dagger}\widetilde{Q} - M_L^2\widetilde{L}^{\dagger}\widetilde{L}
		  - M_U^2\widetilde{U}^*\widetilde{U} - M_D^2\widetilde{D}^*\widetilde{D} 
		  - M_E^2\widetilde{E}^*\widetilde{E} \\[1ex]
		& - \big( A_U \widetilde{U}^*H_u\widetilde{Q} + A_D\widetilde{D}^*H_d\widetilde{Q}
			+ A_E \widetilde{E}^*H_d\widetilde{L} + \rm{h.c.} \big) \\[1ex]
		& - m_{H_u}^2 H_u^*H_u - m_{H_d}^2 H_d^*H_d - \big( b H_u^*H_d+{\rm h.c.} \big) \,.
	\end{split}
	\label{Eq:Lagrangian}
\end{align}

While the soft mass and trilinear parameters appearing in Eq.\ \eqref{Eq:Lagrangian} are assumed to be diagonal matrices in flavour space within the MFV framework, they may comprise non-diagonal entries when relaxing this hypothesis and considering a NMFV scenario. It should be noted that generic SUSY models do not possess any symmetry preventing large off-diagonal elements in soft-SUSY parameters. The soft mass matrices are defined in the Super-CKM (SCKM) basis\footnote{The super-CKM basis is the one in which the up- and down-type quark Yukawa couplings are diagonal matrices.} as:
\begin{gather}
	\begin{split}
	\renewcommand*{\arraystretch}{1.2}
	M^2_Q &= 
	\begin{pmatrix}
		(M_Q)_{11}^2 & (\Delta^Q_{12})^2 & (\Delta^Q_{13})^2 \\
		\boldsymbol{\cdot} & (M_Q)_{22}^2 & (\Delta^Q_{23})^2 \\
		\boldsymbol{\cdot} & \boldsymbol{\cdot} & (M_Q)_{33}^2
	\end{pmatrix}
	\,, \\[1ex]
	\renewcommand*{\arraystretch}{1.2}
	M^2_U &= 
		\begin{pmatrix}
			(M_U)_{11}^2 & (\Delta^U_{12})^2 & (\Delta^U_{13})^2 \\
			\boldsymbol{\cdot} & (M_U)_{22}^2 & (\Delta^U_{23})^2 \\
			\boldsymbol{\cdot} & \boldsymbol{\cdot} & (M_U)_{33}^2
		\end{pmatrix}
	\,, \qquad
	M^2_D = 
		\begin{pmatrix}
			(M_D)_{11}^2 & (\Delta^D_{12})^2 & (\Delta^D_{13})^2 \\
			\boldsymbol{\cdot} & (M_D)_{22}^2 & (\Delta^D_{23})^2 \\
			\boldsymbol{\cdot} & \boldsymbol{\cdot} & (M_D)_{33}^2
		\end{pmatrix}
	\,, \\[1ex]
	\renewcommand*{\arraystretch}{1.2}
	M^2_L &= 
		\begin{pmatrix}
			(M_L)_{11}^2 & (\Delta^L_{12})^2 & (\Delta^L_{13})^2 \\
			\boldsymbol{\cdot} & (M_L)_{22}^2 & (\Delta^L_{23})^2 \\
			\boldsymbol{\cdot} & \boldsymbol{\cdot} & (M_L)_{33}^2
		\end{pmatrix}
	\,, \qquad
	M^2_E = 
		\begin{pmatrix}
			(M_E)_{11}^2 & (\Delta^E_{12})^2 & (\Delta^E_{13})^2 \\
			\boldsymbol{\cdot} & (M_E)_{22}^2 & (\Delta^E_{23})^2 \\
			\boldsymbol{\cdot} & \boldsymbol{\cdot} & (M_E)_{33}^2
		\end{pmatrix} \,
	\end{split}
	\label{Eq:Full_MSSM_soft_matrices}
\end{gather}
which are associated to the left-handed squarks, right-handed up- and down-type squarks, left-handed sleptons and sneutrinos, and right-handed sleptons, respectively. In addition, there are the trilinear coupling matrices:
\begin{gather}
	\begin{split}
	\renewcommand*{\arraystretch}{1.2}
	A_U &= 
		\begin{pmatrix}
			(A_U)_{11} & \Delta^{AU}_{12} & \Delta^{AU}_{13} \\
			\Delta^{AU}_{21} & (A_U)_{22} & \Delta^{AU}_{23} \\
			\Delta^{AU}_{31} & \Delta^{AU}_{32} & (A_U)_{33}
		\end{pmatrix}	
	\,, \quad
	A_D = 
		\begin{pmatrix}
			(A_D)_{11} & \Delta^{AD}_{12} & \Delta^{AD}_{13} \\
			\Delta^{AD}_{21} & (A_D)_{22} & \Delta^{AD}_{23} \\
			\Delta^{AD}_{31} & \Delta^{AD}_{32} & (A_D)_{33}
		\end{pmatrix}		
	\,, \\[1ex]
	\renewcommand*{\arraystretch}{1.2}
	A_E &= 
		\begin{pmatrix}
			(A_E)_{11} & \Delta^{AE}_{12} & \Delta^{AE}_{13} \\
			\Delta^{AE}_{21} & (A_E)_{22} & \Delta^{AE}_{23} \\
			\Delta^{AE}_{31} & \Delta^{AE}_{32} & (A_E)_{33}
		\end{pmatrix} \,
	\label{Eq:Full_MSSM_trilinear}
	\end{split}
\end{gather}
for the up- and down-type squarks and the sleptons. Detailed expressions for the diagonal elements of the matrices given in Eqs.\ \eqref{Eq:Full_MSSM_soft_matrices} and \eqref{Eq:Full_MSSM_trilinear} can be found in Ref.\ \cite{SUSYPrimer}. Note that the soft mass matrices in Eq.\ \eqref{Eq:Full_MSSM_soft_matrices} are symmetric due to the requirement for hermiticity. 

It is convenient to parametrize the off-diagonal, i.e.\ flavour violating, elements of the above matrices in a dimensionless manner by normalizing them to the respective diagonal entries of the sfermion mass matrices. In the SCKM basis, this leads to the following parameters \cite{Ciuchini:2007ha};
\begin{gather}
		(\delta^Q_{LL})_{ij} = \frac{(\Delta^Q_{ij})^2}{(M_Q)_{ii}(M_Q)_{jj}},\quad(\delta^U_{RR})_{ij} = \frac{(\Delta^U_{ij})^2}{(M_U)_{ii}(M_U)_{jj}} \,,\quad
		(\delta^D_{RR})_{ij} = \frac{(\Delta^D_{ij})^2}{(M_D)_{ii}(M_D)_{jj}} \,,\nonumber\\[1ex]
		(\delta^U_{RL})_{ij}=\frac{v_u}{\sqrt{2}}\frac{\Delta^{AU}_{ij}}{(M_Q)_{ii}(M_U)_{jj}} \,,\quad
		(\delta^D_{RL})_{ij} = \frac{v_d}{\sqrt{2}}\frac{\Delta^{AD}_{ij}}{(M_Q)_{ii}(M_D)_{jj}} \,, \label{Eq:MSSM_small_deltas}\\[1ex]
		(\delta^L_{LL})_{ij} = \frac{(\Delta^L_{ij})^2}{(M_L)_{ii}(M_L)_{jj}},\quad (\delta^E_{RR})_{ij} = \frac{(\Delta^E_{ij})^2}{(M_E)_{ii}(M_E)_{jj}} \,,\quad
		(\delta^E_{RL})_{ij} = \frac{v_d}{\sqrt{2}}\frac{\Delta^{AE}_{ij}}{(M_L)_{ii}(M_E)_{jj}} \,,\nonumber
\end{gather}
with $v_u$ and $v_d$ being the vacuum expectation values of the up- and down-type Higgs doublets, respectively. Note that these definitions hold at any scale. In the following, the scales of interest will be the GUT and TeV scales. Moreover, the situation where all off-diagonal NMFV parameters defined in Eq.\ \eqref{Eq:MSSM_small_deltas} vanish corresponds to a scenario with quite minimal flavour violation.

\subsection{$SU(5)$ unification}

The MSSM realization under consideration is based on the gauge group $SU(5)$, which is the smallest group containing the SM gauge group, and can accomodate its matter fields in the $F = \overline{\bf 5}$ and $T={\bf 10}$ representations according to
\begin{align}
	F = \overline{\bf 5} = \left( \begin{array}{c} d_r^c \\ d_b^c \\ d_g^c \\ e^- \\ -\nu_e \end{array} \right)_{\!\!L} \,, 
	\qquad
	T = {\bf 10} = \left( \begin{array}{ccccc} 0 & u_g^c & -u_b^c & u_r & d_r \\ . & 0 & u_r^c & u_b & d_b \\ 
		. & . & 0 & u_g & d_g \\ . & . & . & 0 & e^c \\ . & . & . & . & 0 \end{array}\right)_{\!\!L} \,,
	\label{Eqn:SU5_reps}
\end{align}
where $r, b, g$ denote the quark colours, and $c$ denotes $CP$-conjugated fermions. The Higgs doublets $H_u$ and $H_d$, which break the electroweak symmetry, may arise from $SU(5)$ multiplets $H_{\bf 5}$ and $H_{\overline{\bf 5}}$, provided the colour triplet components are heavy. The $SU(5)$ gauge group may be broken by an additional Higgs multiplet in the ${\bf 24}$ representation developing a vacuum expectation value
\begin{align}
	SU(5) ~\rightarrow~ SU(3)_C\times SU(2)_L\times U(1)_Y \,,
\end{align}
where complete SM quark and lepton families $(Q,u^c,d^c,L,e^c)$ fit into the representations as
\begin{align}
	\begin{split}
		F(\overline{\bf 5}) ~&=~ d^c(\overline{\bf 3},{\bf 1},1/3) \oplus L({\bf 1},\overline{\bf 2},-1/2) \,, \\
		T({\bf 10}) ~&=~ u^c(\overline{\bf 3},{\bf 1},-2/3) \oplus  Q({\bf 3},{\bf 2},1/6)\oplus e^c({\bf 1},{\bf 1},1)\,.
	\end{split}
\end{align}

Including the above arguments into a supersymmetric framework, $SU(5)$ symmetry provides relationships between the soft terms belonging to the supermultiplets within a given representation. For the MSSM under consideration here, we can write down the soft-breaking Lagrangian in terms of $SU(5)$ fields:
\begin{align}
	\begin{split}
		\mathcal{L}^{\rm SU(5) MSSM}_{\rm soft} = 
		& - \frac{1}{2} \big( M_1\widetilde{B}\widetilde{B}+M_2\widetilde{W}\widetilde{W} 
		+ M_3\widetilde{g}\widetilde{g} + \rm{h.c.} \big) \\[1ex]
		& - M_F^2\widetilde{F}^{\dagger}\widetilde{F} - M_T^2\widetilde{T}^{\dagger}\widetilde{T}\\[1ex]
		& - \big( A_{TT} \widetilde{T}^*H_u\widetilde{T} + A_{FT}\widetilde{F}^*H_d\widetilde{T} + \rm{h.c.}\big)\\[1ex]
		& - m_{H_u}^2 H_u^*H_u - m_{H_d}^2 H_d^*H_d - \big( b H_u^*H_d+{\rm h.c.} \big) \,.
	\end{split}
	\label{Eqn:SU(5)_soft_breaking_Lagrangian}
\end{align}	
where $\widetilde{F}$ and $\widetilde{T}$ are the superpartner fields of $F$ and $T$ given in  Eq.\ \eqref{Eqn:SU5_reps}. Comparing this with Eq.\ \eqref{Eq:Lagrangian} leads to the relations
\begin{align}
	\begin{split}
	M^2_{Q} ~=~ M^2_{U} ~=~ M^2_{E} ~&\equiv~ M^2_{T} \,,\\
	M^2_{D} ~=~ M^2_{L} ~&\equiv~ M^2_{F} \,,\\
	A_{D} ~=~ (A_{E})^T ~&\equiv~ A_{FT} \,,\\
	A_{U} ~&\equiv~ A_{TT} \,,
	\end{split}
	\label{Eq:matrix_defintions}
\end{align}
that hold at the GUT scale. Note that renormalization group evolution towards lower scales will spoil these relations.

\subsection{The $A_4\times SU(5)$ model}
In addition to the $SU(5)$ grand unification, we impose an $A_4$ (alternating group of order 4) flavour symmetry on the model under consideration. To this end, we unify the three families of $F = {\bf \bar{5}} = (d^c, L)$ into the triplet of $A_4$ leading to a unified soft mass parameter $m_F$ for the three generations\footnote{In principle, any  group that admits triplet representations can give degenerate soft masses here.}. The three families of $T_i = {\bf 10}_i = (Q, u^c, e^c)_i$ are singlets of $A_4$, which means that the three generations may have independent soft mass parameters $m_{T_1}$, $m_{T_2}$, $m_{T_3}$ \cite{Callen:2012kd, Antusch:2013wn, Cooper:2012wf, Cooper:2010ik, Bjorkeroth:2015ora}.

Through breaking the discrete symmetry just below the GUT scale, we can induce flavour violation in our soft parameters. We express this primordial flavour violation as the matrices $M^2_T$, $M^2_F$, $A_{FT}$, and $A_{TT}$ analogously to Eq.\ \eqref{Eq:Full_MSSM_soft_matrices} in the flavour basis of $A_4$, that is, before rotation to the SCKM:
\begin{gather}
	\renewcommand*{\arraystretch}{1.5}
	\begin{split}
		M^2_T = \begin{pmatrix}
			m_{T_1}^2 & (\Delta^T_{12})^2 & (\Delta^T_{13})^2 \\
			\boldsymbol{\cdot} & m_{T_2}^2 & (\Delta^T_{23})^2 \\
			\boldsymbol{\cdot} & \boldsymbol{\cdot} & m_{T_3}^2
		\end{pmatrix}\,, \qquad
		M^2_F = \begin{pmatrix}
			m_{F}^2 & (\Delta^F_{12})^2 & (\Delta^F_{13})^2 \\
			\boldsymbol{\cdot} & m_{F}^2 & (\Delta^F_{23})^2 \\
			\boldsymbol{\cdot} & \boldsymbol{\cdot} & m_{F}^2
		\end{pmatrix}\,, \\[3ex]
		A_{FT} = \begin{pmatrix}
			(A_{FT})_{11} & \Delta^{FT}_{12} & \Delta^{FT}_{13} \\
			\Delta^{FT}_{21} & (A_{FT})_{22} & \Delta^{FT}_{23} \\
			\Delta^{FT}_{31} & \Delta^{FT}_{32} & (A_{FT})_{33}
		\end{pmatrix}\,, \qquad
		A_{TT} = \begin{pmatrix}
			(A_{TT})_{11} & \Delta^{TT}_{12} & \Delta^{TT}_{13} \\
			\Delta^{TT}_{21} & (A_{TT})_{22} & \Delta^{TT}_{23} \\
			\Delta^{TT}_{31} & \Delta^{TT}_{32} & (A_{TT})_{33}
		\end{pmatrix}\
	\end{split}
	\label{Eq:SU(5)_dimensionful_NMFV_matrices}
\end{gather}

Note that the breaking of $A_4$ enforces off-diagonal elements of the $M_T^2$ and $A_{FT}$ matrices in Eq.\ \eqref{Eq:SU(5)_dimensionful_NMFV_matrices} to be smaller than diagonal entries, and we also assume that off-diagonal elements in the other matrices are small\footnote{This assumption becomes inevitable when one considers an additional U(1) symmetry as per Ref. \cite{A4SU5}, which is required for the Froggatt-Nielsen mechanism and to supply correct flavon vev alignment.}. This provides a theoretical motivation for small-but-non-zero flavour violation in such a class of models. $SU(5)$ gives the following relationships between the dimensionless NMFV parameters in the basis before rotation to the SCKM (as denoted by the subscript `0'):
\begin{align}
	\begin{split}
		\delta^{Q_0}_{LL} ~=~ \delta^{U_0}_{RR} ~=~  \delta^{E_0}_{RR} ~&\equiv~ \deltat, \\	
		\quad\delta^{D_0}_{RR} ~=~ \delta^{L_0}_{LL} ~&\equiv~ \deltaf \,, \\
		\delta^{D_0}_{RL} ~=~ (\delta^{E_0}_{RL})^T ~&\equiv~ \deltaft \,,\\
		\delta^{U_0}_{RL} ~&\equiv~ \deltatt \,
	\label{Eq:NMFV_GUT_A4xSU5_deltas}
	\end{split}
\end{align}
These four matrices parameterise the flavour violation in the $A_4\times SU(5)$ setup studied here. Note that $\deltat$, $\deltaf$ and $\deltatt$ are necessarily symmetric whereas $\deltaft$ is not (see Eqs.\ \eqref{Eqn:SU(5)_soft_breaking_Lagrangian} and \eqref{Eq:SU(5)_dimensionful_NMFV_matrices}) leading to a total of 15 NMFV parameters at the GUT scale.

It is apparent that we have flavour violation at phenomenological scales from two distinct sources: The presence of off-diagonal elements in various coupling matrices at the GUT scale due to $A_4$ breaking, and further effects on the off-diagonal elements induced by RGE running. We do not consider a specific breaking mechanism or pattern for the discrete symmetry.


\section{Setup and tools}
\label{Sec:Setup}

The aim of the present study is to assess the impact of flavour violating parameters introduced at the GUT scale on low-energy physics. In order to work with a concrete framework, we focus on the model based on $A_4 \times SU(5)$ presented in Sec.\ \ref{Sec:Model}. We test the high-scale model against the Dark Matter (DM) relic density along with leptonic and hadronic flavour changing observables and the mass of the Higgs boson.

\subsection{MFV reference points}
\label{sec:MFV_benchmark_points}

In order to focus on the impact of NMFV terms in the Lagrangian of our model, we start by choosing suitable reference scenarios respecting the MFV paradigm. From previous work \cite{A4SU5} it is apparent that successfully imposing the dark matter relic density as well as the anomalous magnetic moment of the muon on the $A_4 \times SU(5)$ framework requires rather specific parameter configurations. More precisely, the corresponding parameter points feature a physical spectrum where the ``right-handed'' smuon is light and almost mass-degenerate with the lightest neutralino, which is bino-like. This allows to simultaneously satisfy the $(g-2)_{\mu}$ and relic density constraints \cite{PDG2016, Planck2016}. For our study, we choose two MFV reference scenarios, which are summarized in Table \ref{tab:RefScenarioGUT}.

\begin{table}
	\centering
	\renewcommand{\arraystretch}{1.2}
	\begin{tabular}{|c|c|c|c|}
		\hline
		 & Parameter/Observable & Scenario 1 & Scenario 2 \\
		\hline
		\hline
		\multirow{11}{*}{\rotatebox[origin=c]{90}{MFV Parameters at GUT scale}}
		& $m_F$ & 5000 & 5000 \\ 
		& $m_{T_1}$ & 5000 & 5000 \\
		& $m_{T_2}$ & 200 & 233.2 \\
		& $m_{T_3}$ & 2995 & 2995 \\
		\cline{2-4}
		& $(A_{TT})_{33}$ & -940 & -940 \\
		& $(A_{FT})_{33}$ & -1966 & -1966 \\
		\cline{2-4}
		& $M_1$ & 250.0 & 600.0 \\
		& $M_2$ & 415.2 & 415.2 \\
		& $M_3$ & 2551.6 & 2551.6 \\
		\cline{2-4}
		& $m_{H_u}$ & 4242.6 & 4242.6 \\
		& $m_{H_d}$ & 4242.6 & 4242.6 \\
		\hline
		& $\tan\beta$ & 30 & 30 \\
		& $\mu$ & -2163.1 & -2246.8 \\
		\hline
		\hline
		\multirow{10}{*}{\rotatebox[origin=c]{90}{Physical masses}}
		& $m_h$ & 126.7 & 127.3 \\
		& $m_{\widetilde{g}}$ & 5570.5 & 5625.7 \\
		& $m_{\widetilde{\mu}_L}$ & 4996.7 & 4997.5 \\
		& $m_{\widetilde{\mu}_R}$ & 102.1 & 254.4 \\
		& $m_{\widetilde{\chi}_1^0}$ & 94.6 & 250.4 \\
		& $m_{\widetilde{\chi}_2^0}$ & 323.6 & 322.0 \\
		& $m_{\widetilde{\chi}_3^0}$ & 2248.8 & 2331.1 \\
		& $m_{\widetilde{\chi}_4^0}$ & 2248.8 & 2331.2\\
		& $m_{\widetilde{\chi}_1^\pm}$ & 323.8 & 322.2 \\
		& $m_{\widetilde{\chi}_2^\pm}$ & 2249.8 & 2332.2 \\
		\hline
		& $\Omega_{\widetilde{\chi}^0_1}h^2$ & 0.116 & 0.120 \\
		& $\sigma^{\rm proton}_{\rm SI}/10^{-14}\,{\rm pb}$ & 2.987 & 1.055 \\
		& $\sigma^{\rm neutron}_{\rm SI}/10^{-14}\,{\rm pb}$ & 3.249 & 0.986 \\
		\hline
	\end{tabular}
	\caption{GUT scale inputs together with selected physical masses and relevant TeV scale parameters for the two MFV reference scenarios. First and second generation trilinear couplings are set to zero. Further squark and slepton masses which are beyond the reach of current experiments are not shown. Unless otherwise illustrated, dimensionful quantities are given in GeV. DM direct detection cross-sections are given for both protons and neutrons.}
	\label{tab:RefScenarioGUT}
\end{table}
The first reference point of our choice corresponds to the scenario labelled `BP4' in Ref.\ \cite{A4SU5}. For practical reasons, mainly due to including NMFV terms at the GUT scale, we do not make use of the same version of the spectrum generator {\tt SPheno}. In consequence, effects from renormalization group running differ slightly, and we have adapted the input parameters of the original BP4 reference scenario to the ones given in Table \ref{tab:RefScenarioGUT}. However, note that, although there is a small deviation for the TeV scale parameters as compared to scenario BP4 of Ref.\ \cite{A4SU5}, the phenomenological aspects of our reference scenario at the TeV scale are unaffected. Let us recall that the rather low smuon mass parameter, $m_{T_2} = 200$ GeV, which leads to the physical mass $m_{\tilde{\mu}_R} = 102.1$ GeV, is required in order to satisfy simultaneously the $(g-2)_{\mu}$ and relic density constraints as discussed in Ref.\ \cite{A4SU5}. This particular choice for $m_{T_2}$ is an assumption in this work.

While current limits on ``right-handed'' smuons still allow masses as low as about 100 GeV \cite{ATLAS2014Sleptons}, this first scenario is going to be severely challenged by ongoing LHC searches. For this reason, we choose to include a second reference point which is inspired by the first one but features larger smuon and neutralino masses. This still allows satisfaction of the relic density constraint due to efficient co-annihilation and avoids LHC limits to be published in the near future. Note that, however, the higher smuon mass $m_{\tilde{\mu}_R} \sim 250$ GeV does not resolve the tension between the Standard Model and the experimental value of $(g-2)_{\mu}$. Let us emphasize that both reference scenarios capture the essential results of Ref.\ \cite{A4SU5}, namely almost mass-degenerate ``right-handed'' smuon and bino-like neutralino, while all other MSSM states are essentially decoupled. 

In both reference scenarios, the required neutralino relic density is met thanks to efficient co-annihilation with the smuon and even smuon pair annihilation. All (co)annihilation contributions are summarized in Table \ref{tab:MFV_OmegaDM_relevant_decay_modes}. Neutralino pair annihilation mainly proceeds through $t$- and $u$-channel smuon exchange, while smuon pair annihilation proceeds through neutralino $t$- or $u$-channel exchange. Moreover, the relative importance of the co-annihilation and smuon pair annihilation with respect to the neutralino pair annihilation is governed by the Boltzmann factor involving the mass difference of the two particles \cite{Gondolo:1990dk}. The smuon mass therefore plays a central role in this context. Considering NMFV, the off-diagonal elements of the matrices in Eqs.\ \eqref{Eq:Full_MSSM_soft_matrices} and \eqref{Eq:Full_MSSM_trilinear} not only violate flavour but can in addition have a significant impact on the smuon mass and thus on the relic density.

\begin{table}
	\centering
	\renewcommand{\arraystretch}{1.2}
	\begin{tabular}{|c|p{2.5cm}|p{2.5cm}|}
		\hline
		\multicolumn{1}{|c|}{\multirow{2}{*}{Annihilation channel}} & \multicolumn{2}{c|}{Relative contribution to $\Omega_{\tilde{\chi}^0_1}h^2$}\\
		\cline{2-3}
		& \hfil Scenario 1 & \hfil Scenario 2 \\
		\hline
		$\tilde{\chi}^0_1 \, \tilde{\chi}^0_1 \rightarrow \mu \, \bar{\mu}$ & \hfil$27\%$ & \hfil$2\%$ \\
		$\tilde{\chi}^0_1 \, \tilde{\mu}_R \rightarrow \mu \, \gamma$ & \hfil$45\%$  & \hfil$31\%$\\
		$\tilde{\chi}^0_1 \, \tilde{\mu}_R \rightarrow \mu \, Z^0$ & \hfil$8\%$ & \hfil$8\%$ \\
		$\tilde{\mu}_R \, \tilde{\mu}_R \rightarrow \mu \, \mu$ & \hfil$10\%$ & \hfil$37\%$ \\		
		$\tilde{\mu}_R \, \tilde{\mu}_R^* \rightarrow \gamma \, \gamma$ & \hfil$3\%$ & \hfil$11\%$ \\
		\hline
	\end{tabular}
	\caption{Dominant annihilation channels contributing to the annihilation cross-section and the neutralino relic density in the two MFV reference scenarios of Table \ref{tab:RefScenarioGUT}.}
	\label{tab:MFV_OmegaDM_relevant_decay_modes}
\end{table}

\subsection{Introducing NMFV}

Starting from the two MFV reference points, we study the impact of flavour violating soft terms by perturbing around this scenario. Keeping the MFV parameters fixed at the values given in Table \ref{tab:RefScenarioGUT}, we perform a random scan on the flavour violating parameters introduced in Eq.\ \eqref{Eq:NMFV_GUT_A4xSU5_deltas} at the GUT scale using flat prior distributions. In practice, we vary the NMFV parameters both independently and as part of a multi-dimensional scan over all parameters simultaneously. We subsequently study the impact of the constraints detailed in Table \ref{Tab:Constraints}.

\begin{table}
	\centering
	\renewcommand{\arraystretch}{1.2}
	\begin{tabular}{|c|c|c|c|}
		\hline
		Observable & Constraint & Remarks & Refs. \\
		\hline
		$m_h$ & $\left( 125.2 \pm 2.5 \right)$ GeV & (SPheno th.) & \cite{PDG2018, SPheno2003, SPheno2011} \\
		\hline
		$\mathrm{BR}(\mu \rightarrow e\gamma)$ & $ <4.2 \times 10^{-13}$ & 90\% (exp.)  & \cite{PDG2018} \\
		$\mathrm{BR}(\mu \rightarrow 3e)$ & $ <1.0 \times 10^{-12}$ & 90\% (exp.)  & \cite{PDG2018} \\
		$\mathrm{BR}(\tau \rightarrow e\gamma)$ & $ <3.3 \times 10^{-8}$ & 90\% (exp.)  & \cite{PDG2018} \\
		$\mathrm{BR}(\tau \rightarrow \mu\gamma)$ & $ <4.4 \times 10^{-8}$ & 90\% (exp.)  & \cite{PDG2018} \\
		$\mathrm{BR}(\tau \rightarrow 3e)$ & $ <2.7 \times 10^{-8}$ & 90\% (exp.)  & \cite{PDG2018} \\
		$\mathrm{BR}(\tau \rightarrow 3\mu)$ & $ <2.1 \times 10^{-8}$ & 90\% (exp.)  & \cite{PDG2018} \\
		$\mathrm{BR}(\tau \rightarrow e^-\mu\mu)$ & $ <2.7 \times 10^{-8}$ & 90\% (exp.)  & \cite{PDG2018} \\
		$\mathrm{BR}(\tau \rightarrow e^+ \mu \mu)$ & $ <1.7 \times 10^{-8}$ & 90\% (exp.)  & \cite{PDG2018} \\
		$\mathrm{BR}(\tau \rightarrow \mu^- e e)$ & $ <1.8 \times 10^{-8}$ & 90\% (exp.)  & \cite{PDG2018} \\
		$\mathrm{BR}(\tau \rightarrow \mu^+ e e)$ & $ <1.5 \times 10^{-8}$ & 90\%  (exp.) & \cite{PDG2018} \\
		\hline
		$\mathrm{BR}(B \rightarrow X_s \gamma)$ & $\left( 3.32 \pm 0.18 \right) \times 10^{-4}$ & $2\sigma$ (exp.) & \cite{HFLAF2017}\\
		$\mathrm{BR}(B_s \rightarrow \mu \mu)$ & $\left( 2.7 \pm 1.2 \right) \times 10^{-9}$ & $2\sigma$ (exp.) &\cite{PDG2018}\\
		$\Delta M_{B_s}$ & $\left( 17.757 \pm 0.042 \pm 2.7 \right)$ ps$^{-1}$ & $2\sigma$ (exp.), (th.)  &\cite{PDG2018, SMDelMBs} \\
		$\Delta M_K$ & $\left( 3.1 \pm 1.2 \right) \times 10^{-15} $ GeV & $2\sigma$ (th.) & \cite{PDG2018,SMDelMK}\\
		$\epsilon_K$ & $2.228 \pm 0.29$ & $2\sigma$ (th.)&\cite{PDG2018, SMDelMK}\\
		\hline
		$\Omega_{\rm CDM}h^2$ & $0.1198 \pm 0.0042$ & $2\sigma$ (exp.), 1\% (th.) & \cite{Planck2016, micrOMEGAs2001, micrOMEGAs2004, micrOMEGAs2016} \\
		\hline
	\end{tabular}
	\caption{Experimental constraints imposed on the $A_4 \times SU(5)$ MSSM parameter space in our study. Upper limits are given at the 90\% confidence level, while two-sided limits are understood at the $2\sigma$ confidence level.}
	\label{Tab:Constraints}
\end{table}

More precisely, we require the Higgs-boson mass to be reasonably close to the observed value of about 125 GeV, where we account for a theory uncertainty of 2.5 GeV from the {\tt SPheno} calculation. For the $B_s$-meson oscillation, we consider the experimental value $\Delta M_{B_s} = \left( 17.757 \pm 0.021 \right)$ ps$^{-1}$ \cite{PDG2016} and add a theory uncertainty of 1.35 ps$^{-1}$ \cite{SMDelMBs} which dominates over the experimental error. For the neutralino relic density, we require that the lightest neutralino accounts for the totality of observed cold dark matter. The error given by the Planck collaboration is augmented in order to take into account the 1\% accuracy of the theoretical calculation of the relic density by {\tt micrOMEGAs}.
For further details on experimental constraints we refer the reader to Table \ref{Tab:Constraints} and the references therein.

Finally, note that although the reference scenarios defined in Table \ref{tab:RefScenarioGUT} have in part been obtained considering the anomalous magnetic moment of the muon as a key observable \cite{A4SU5}, we do not take into account this constraint here. Since $(g-2)_{\mu}$ is a flavour-conserving process, we do not expect sizeable effects from NMFV terms on this observable within the ranges that are allowed from the other constraints. 

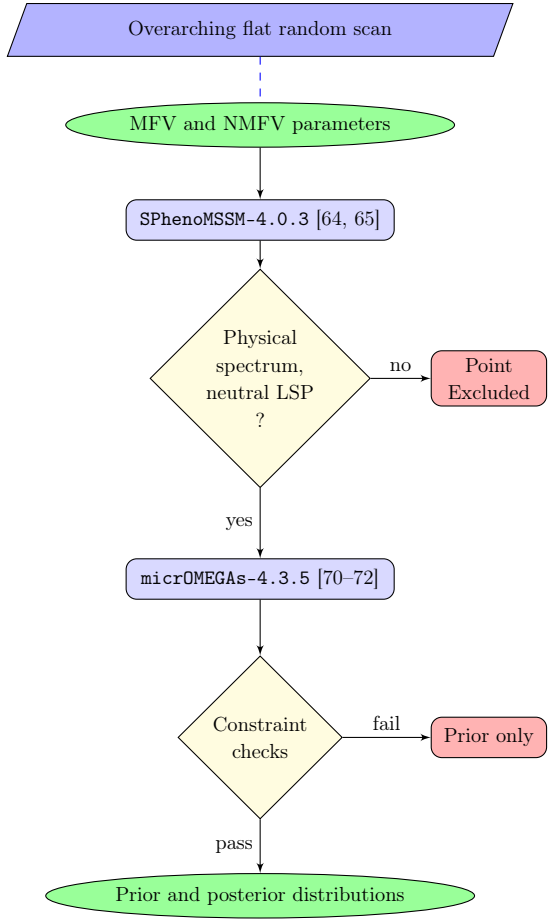
\begin{figure}
	\centering
	\begin{NoHyper}
	\begin{tikzpicture}[node distance = 1.8cm]
		\begin{scope}[scale=0.7, transform shape]
			\node [cloud1] (init) {MFV and NMFV parameters};
			\node [block,below of=init] (mats) {{\tt SPhenoMSSM-4.0.3} \cite{SPheno2003, SPheno2011}};
			\node [decision,below of=mats] (phys) {Physical spectrum, neutral LSP\\?};ese two points allow us
			\node [block0,right of=phys, xshift=2.5cm] (exclude1) {Point Excluded};
			\node [block,below of=phys,yshift=-2.0cm] (micromegas) {{\tt micrOMEGAs-4.3.5} \cite{micrOMEGAs2001, micrOMEGAs2004, micrOMEGAs2016}};
			\node [decision,below of=micromegas] (constcheck) {Constraint checks};
			\node [block0,right of=constcheck,xshift=2.5cm] (exclude2) {Prior only};
			\node [cloud1, below of=constcheck] (predictions) {Prior and posterior distributions};
			\node [io, above of=init] (rand scan) {Overarching flat random scan};
			
			\node [left of=rand scan] (left1) {};
			\node [left of=init] (left2) {} ; 
			
			\path [line] (init) -- (mats);
			\path [line] (mats) -- (phys);
			\path [line] (micromegas) -- (constcheck);
			\draw [dashed, blue] (rand scan) -- (init);
			
			\draw [line] (phys) -- node [anchor=south] {no} (exclude1);
			\draw [line] (phys) -- node [anchor=east] {yes} (micromegas);
			\draw [line] (constcheck) -- node [anchor=south] {fail} (exclude2);
			\draw [line] (constcheck) -- node [anchor=east] {pass} (predictions);
		\end{scope}
	\end{tikzpicture}
	\caption{Illustration of the computational procedure applied to each individual point of our parameter scan.}
	\label{Fig:Flowchart}
	\end{NoHyper}
\end{figure}

For numerical evaluation, we make use of the spectrum generator {\tt SPheno~4.0.3} \cite{SPheno2003, SPheno2011}, where we have included the MSSM with general flavour mixing using the {\tt Mathematica} package {\tt SARAH~4.12.3} \cite{SARAH2009, SARAH2010, SARAH2013, SARAH2014}. From the resulting code {\tt SPhenoMSSM} we obtain through two-loop renormalization group equations for the soft-breaking parameters and the physical mass spectrum at the TeV scale, as well as numerical predictions for flavour observables listed in Table \ref{Tab:Constraints}. The neutralino relic density $\Omega_{\tilde{\chi}^0_1}h^2$ is computed using the public package {\tt micrOMEGAs~4.3.5} \cite{micrOMEGAs2001, micrOMEGAs2004, micrOMEGAs2016}. Again, we have used {\tt SARAH} to obtain the {\tt CalcHEP} model files necessary to accomodate NFMV effects in the calculation. Our computational setup is summarized in Fig.\ \ref{Fig:Flowchart}. The mass spectrum obtained from {\tt SPhenoMSSM} is handed to {\tt micrOMEGAs} by making use of the {\it SUSY Les Houches Accord 2} \cite{SLHA2}. Note that, since the spin-independent scattering cross-sections related to direct dark matter detection given in Table \ref{tab:RefScenarioGUT} are relatively low as compared to the corresponding experimental limits, we do not explicitly evaluate these cross-section in our NMFV scan. 

Before running {\tt SPheno}, we first need to perform a CKM transformation to certain GUT scale matrices to comply with the basis that {\tt SPheno} requires for the input parameters (see Appendix \ref{Appendix:SCKM}). Let us note that, for typical values of Yukawa parameters inserted into our MFV reference points, CKM matrix running between the GUT and TeV scales has been found to be negligible. We therefore assume that the CKM matrix is identical across all scales.

In the full multi-dimensional scan, the studied range for each parameter is set empirically to give reasonable computational efficiency as informed by one-dimensional scans over individual parameters. The obtained ranges have been increased slightly to be able to study whether correlations between the different NMFV parameters may result in larger allowed ranges as compared to the one-dimensional scan. The applied limiting values for each MFV scenario under consideration and for each NMFV parameter are given in Table \ref{Tab:ParameterRange}.

\begin{table}
	\begin{center}
	\renewcommand{\arraystretch}{1.3}
	\begin{tabular}{|c|c|c|}
		\hline
		Parameters & Scenario 1 & Scenario 2 \\
		\hline
		\hline
		$(\deltat)_{12}$ & $[-2.00,2.00]\times 10^{-2}$ & $[-5.57,5.15]\times10^{-2}$ \\
		$(\deltat)_{13}$ & $[-8.01,8.01]\times 10^{-2}$ & $[-0.267,0.301]$ \\
		$(\deltat)_{23}$ & $0.0$ & $[-5.73,5.73]\times10^{-2}$ \\
		\hline
		$(\deltaf)_{12}$ & $[-8.00,8.00]\times 10^{-3}$ & $[-8.00,8.00]\times10^{-3}$\\
		$(\deltaf)_{13}$ & $[-1.00,1.00]\times 10^{-2}$ & $[-8.00,8.00]\times10^{-2}$\\
		$(\deltaf)_{23}$ & $[-1.60,1.60]\times 10^{-2}$ & $[-8.00,8.00]\times10^{-2}$\\
		\hline
		\hline
		$(\deltatt)_{12}$ & $[-8.69,10.43]\times10^{-4}$ & $[-7.46,8.95]\times10^{-4}$\\
		$(\deltatt)_{13}$ & $[-1.74, 1.74]\times10^{-3}$ & $[-3.48,1.74]\times10^{-3}$\\
		$(\deltatt)_{23}$ & $[-0.0174,0.145]$ & $[-0.0871,0.124]$\\
		\hline
		$(\deltaft)_{12}$ & $[-4.64,4.64]\times10^{-5}$ & $[-5.47,5.47]\times10^{-5}$\\
		$(\deltaft)_{13}$ & $[-7.74,7.74]\times10^{-5}$ & $[-3.87,3.87]\times10^{-4}$\\
		$(\deltaft)_{21}$ & $0.0$ & $[-1.04,1.04]\times10^{-4}$\\
		$(\deltaft)_{23}$ & $[-1.16,1.16]\times10^{-4}$ & $[-2.32,2.32]\times10^{-4}$\\
		$(\deltaft)_{31}$ & $[-1.39,1.39]\times10^{-5}$ & $[-8.81,8.81]\times10^{-5}$\\
		$(\deltaft)_{32}$ & $0.0$ & $[-1.49,1.49]\times10^{-4}$\\
		\hline
	\end{tabular}
	\end{center}
	\caption{Ranges of the NMFV parameters defined at the GUT scale (see Eq.\ \eqref{Eq:NMFV_GUT_A4xSU5_deltas}) for our multi-dimensional scans around the reference scenarios. Those parameters given as $0.0$ have been switched off, since even small variations lead to tachyonic mass spectra and/or a charged LSP.}
	\label{Tab:ParameterRange}
\end{table}

Already from the individual scans, it becomes apparent that for certain NMFV parameters, especially in the case of Scenario 1, small deviations from the MFV case can induce either a charged dark matter candidate (the smuon in this case) or tachyonic mass spectra. We therefore set 
\begin{align}
	(\deltat)_{23} ~=~ (\deltaft)_{21} ~=~ (\deltaft)_{32} ~=~ 0
\end{align}
throughout the analysis of Scenario 1, and scan over the remaining 12 NMFV parameters according to Table \ref{Tab:ParameterRange}. This situation does not occur for Scenario 2, where we vary all 15 NMFV parameters.

Starting from parameters at the GUT scale, we test each point against the observables listed in Table \ref{Tab:Constraints}. Points which do not satisfy all the imposed constraints within the associated uncertainties are collected in the prior distribution only, while those which comply with all constraints are in addition recorded as part of the posterior distribution. In examining the latter, we obtain the allowed ranges for each of the NMFV parameters. In addition, by comparing the prior and posterior distributions, and taking into account posterior distributions based on a single constraint, we identify the most important constraints among those listed in Table \ref{Tab:Constraints} for each NMFV parameter. The results are presented in the next Section.


\section{Results and Discussion}
\label{Sec:Results}

In this Section we present the results of our analysis. Before coming to a more detailed discussion, we start by presenting the general aspects and the obtained limits on the NMFV parameters, presented in Table \ref{Tab:constrained_parameters_range}. Ultimately, we perform two different kinds of scan on the parameter space: ``individual'' scans, where only a single $\delta$ is varied and all others are set to zero, and ``simultaneous'' scans where all of the NMFV parameters are varied at the same time according to the ranges given in Table \ref{Tab:ParameterRange}. 

\begin{table}
	\centering
	\renewcommand{\arraystretch}{1.3}
	\scalebox{0.82}{
	\begin{tabular}{|c||c|c||c|c|}
		\hline
		Parameters & Scenario 1  & Most constraining obs. 1 & Scenario 2 & Most constraining obs. 2\\
		\hline
		$(\delta^{T})_{12}$  &    [-0.015, 0.015] & $\mu \rightarrow 3e$, $\mu \rightarrow e \gamma$, $\Omega_{\tilde{\chi}_{1}^0 }h^2$ & [-0.12, 0.12]$^\dagger$  & $\Omega_{\tilde{\chi}_{1}^0 }h^2$, $\mu \rightarrow e \gamma$ \\
		$(\delta^{T})_{13}$ & ]-0.06, 0.06[   & $ \Omega_{\tilde{\chi}_{1}^0 }h^2$ &  [-0.3, 0.3]$^\dagger$ & $\Omega_{\tilde{\chi}_{1}^0 }h^2$  \\
		$(\delta^{T})_{23}$ & [0,0]* &  $ \Omega_{\tilde{\chi}_{1}^0 }h^2$, $\mu \rightarrow 3e$, $\mu \rightarrow e \gamma$ & [-0.1, 0.1]$^\dagger$ & $\Omega_{\tilde{\chi}_{1}^0 }h^2$, $\mu \rightarrow 3e$, $\mu \rightarrow e \gamma$, \\
		\hline
		$(\delta^{F})_{12}$ & [-0.008, 0.008]  & $\mu \rightarrow 3e$, $\mu \rightarrow e \gamma$ &  [-0.015, 0.015]$^\dagger$ & $\mu \rightarrow 3e$, $\mu \rightarrow e \gamma$ \\
		$(\delta^{F})_{13}$ & ]-0.01, 0.01[  & $\mu \rightarrow e \gamma$ & [-0.15, 0.15]$^\dagger$ & $\mu \rightarrow 3e$, $\mu \rightarrow e \gamma$\\
		$(\delta^{F})_{23}$ & ]-0.015, 0.015[  & $\mu \rightarrow e \gamma$, $\Omega_{\tilde{\chi}_{1}^0}h^2$ & [-0.15, 0.15]$^\dagger$ & $\Omega_{\tilde{\chi}_{1}^0}h^2$, $\mu \rightarrow e \gamma$, $\mu \rightarrow 3e$ \\
		\hline
		$(\delta^{TT})_{12}$ & [-3, 3.5] $\times 10^{-5}$  & prior &  [-1, 1.5]$^\dagger$ $\times 10^{-3}$ & prior, $\Omega_{\tilde{\chi}_{1}^0}h^2$ \\
		$(\delta^{TT})_{13}$ & ]-6, 7[ $\times 10^{-5}$  & prior, $\Omega_{\tilde{\chi}_{1}^0}h^2$ & [-4, 2.5]$^\dagger$ $\times 10^{-3}$ & prior, $\Omega_{\tilde{\chi}_{1}^0}h^2$ \\
		$(\delta^{TT})_{23}$ & ]-0.5, 4[ $\times 10^{-5}$ & prior, $\Omega_{\tilde{\chi}_{1}^0}h^2$ & [-0.25, 0.2]$^\dagger$ & prior, $\Omega_{\tilde{\chi}_{1}^0}h^2$\\
		\hline
		$(\delta^{FT})_{12}$ & [-0.0015, 0.0015] &  $\Omega_{\tilde{\chi}_{1}^0 }h^2$ & [-1.2, 1.2]$^\dagger$ $\times 10^{-4}$ & $\mu \rightarrow 3e$, $\Omega_{\tilde{\chi}_{1}^0 }h^2$,  $\mu \rightarrow e \gamma$ \\
		$(\delta^{FT})_{13}$ & ]-0.002, 0.002[  & $\Omega_{\tilde{\chi}_{1}^0 }h^2$ & [-5, 5] $\times 10^{-4}$ & $\Omega_{\tilde{\chi}_{1}^0 }h^2$, $\mu \rightarrow 3e$,  $\mu \rightarrow e \gamma$ \\
		$(\delta^{FT})_{21}$ & [0,0]*  & prior & [-1.2, 1.2]$^\dagger$ $\times 10^{-4}$ & $\Omega_{\tilde{\chi}_{1}^0 }h^2$, prior\\
		$(\delta^{FT})_{23}$ & ]-0.0022, 0.0022[  & $\Omega_{\tilde{\chi}_{1}^0}h^2$ & [-6, 6]$^\dagger$ $\times 10^{-4}$ & $\mu \rightarrow 3e$, $\Omega_{\tilde{\chi}_{1}^0 }h^2$,  $\mu \rightarrow e \gamma$\\
		$(\delta^{FT})_{31}$ & ]-0.0004, 0.0004[ & $\Omega_{\tilde{\chi}_{1}^0}h^2$ & [-2, 2]$^\dagger$ $\times 10^{-4}$ & $\Omega_{\tilde{\chi}_{1}^0 }h^2$\\
		$(\delta^{FT})_{32}$ & [0,0]*  & prior &[-1.5, 1.5] $\times 10^{-4}$ & $\Omega_{\tilde{\chi}_{1}^0 }h^2$\\
		\hline
	\end{tabular}
	}
	\caption{Estimated allowed GUT scale flavour violation for both reference scenarios and impactful constraints ordered from the most to the least constraining. Where square brackets are shown open, we scan up to these values but even if we noticed some impact from constraints, it seems that the allowed region can be larger and extrapolation to concrete limits is not straightforward. $^*$ denotes parameters fixed to 0 in order to satisfy LSP and physical mass spectrum requirements. $^\dagger$ stands for extrapolated ranges, meaning that the posterior does not actually drop to 0 but extrapolation to a limit is reasonable. A parameter that is constrained by `prior' is limited by LSP and physical mass requirement.} 
	\label{Tab:constrained_parameters_range}
\end{table}

From the multi-dimensional scan, we conclude that for the majority of the considered NMFV parameters, the most sensitive observables are the branching ratios of $\mu \rightarrow e\gamma$ and $\mu \rightarrow 3e$, as well as the neutralino relic density $\Omega_{\tilde{\chi}^0_1}h^2$. As discussed in Section \ref{Sec:Setup}, the impact of the relic density can be attributed to the small mass difference between the neutralino and the smuon, which depends strongly on the off-diagonal elements in the slepton mass matrix. Since both our reference scenario exhibit a relatively small value of $(m_T)_{22}$, already rather tiny flavour violating elements can be excluded by current data.

Although the experimental limit is more stringent (by about a factor of two) for the decay $\mu \to e\gamma$, the $\mu \to 3e$ decay has about the same constraining power and is in certain cases even the dominant constraint. This is explained as follows: The amplitude of $\mu \to e\gamma$ is helicity-suppressed, and therefore contains a suppression factor $m_e/m_{\mu}$. While this is also the case for $\mu \to 3e$ diagrams related to those of $\mu \to e\gamma$, there are additional four-point diagrams, where the helicity suppression is lifted since no photon is involved. Despite the additional gauge coupling and the greater degree of loop suppression, these diagrams are numerically competitive to those of $\mu \to e\gamma$.

\begin{figure}[H]
	\hspace{1.6cm}
	\Large
	\begin{tikzpicture}
	\begin{feynman}
	\vertex (a) {\(\mu\)};
	\vertex [right=of a] (b);
	\vertex [right=of b] (c); 
	\vertex [right=of c] (d) {\(e\)};
	\diagram* {
		(a) -- [fermion] (b),
		(b) -- [fermion, half left, looseness=1.5] (c),
		(c) -- [scalar, half left, looseness=1.5,insertion=0.5, edge label=$\delta_{12}$] (b),
		(c) -- [fermion] (d),
		(b) -- [boson,half left, looseness=1.5, edge label=$\widetilde{\chi}^0_1$] (c)
	};
	\end{feynman}
	\end{tikzpicture}
	\hspace{1cm}
	\begin{tikzpicture}
	\begin{feynman}
	\vertex (a) {\(\mu\)};
	\vertex [right=of a] (b);
	\vertex [right=of b] (c); 
	\vertex [right=of c] (d) {\(e\)};
	\diagram* {
		(a) -- [fermion] (b),
		(b) -- [fermion, half left, looseness=1.5] (c),
		(c) -- [scalar, half left, looseness=1.5,insertion=0.33, insertion=0.67, edge label=$\,\,\,\delta_{23}\qquad\delta_{13}$] (b),
		(c) -- [fermion] (d),
		(b) -- [boson,half left, looseness=1.5, edge label=$\widetilde{\chi}^0_1$] (c)
	};
	\end{feynman}
	\end{tikzpicture}
	\caption{Feynman diagrams that contribute to $\mu\rightarrow e \gamma$, dashed line represents a slepton and $\delta$ denotes mass insertion parameters. Photon should be taken to be emitted from any particle charged under $QED$.}
	\label{fig:mu_e_gamma_diagrams}
\end{figure}
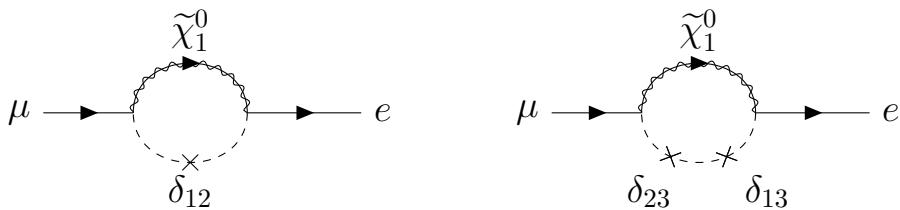

One can see that NMFV parameters mixing the first or second generation with the third generation are also mainly constrained by the decays $\mu \to e\gamma$ and $\mu \to 3e$ rather than by the corresponding $\tau$ decays such as $\tau \to \mu \gamma$ or $\tau \to e\gamma$. This can be traced to the better experimental precision of the muonic decay measurements with respect to the analogous tau decays. Even though NMFV parameters mediating $e-\tau$ or $\mu-\tau$ transitions lead to the dominant contributions of the tau decays, these parameters also can enter into the muon decay amplitudes. For example, if the $\mu \to e\gamma$ process includes a stau in the loop, the corresponding amplitude is proportional to terms including products of the type $(\delta)_{23} (\delta)_{13}$. See Fig.\ \ref{fig:mu_e_gamma_diagrams} for a diagrammatic representation. Since the muon decay limits are stronger than the tau decay limits by four to five orders of magnitude, the $e-\tau$ and $\mu-\tau$ mixing parameters are constrained by the $e-\mu$ processes first. We have explicitly checked this by artificially lowering the bounds on tau decays. In this case, the tau decay becomes the dominant constraint for the $(\delta)_{13}$ and $(\delta)_{23}$ parameters.
Finally, we observe that the constraints coming from the hadronic sector, such as the decays $B \to X_s \gamma$ or $B_s \to \mu^+ \mu^-$, which are dominant in the case of NMFV in the squark sector alone \cite{NMFV2015}, are not competitive as compared to the leptonic constraints mentioned above. This can be traced to the greater experimental precision of dedicated leptonic measurements compared to meson decays.

\subsection{Scan around Scenario 1}

\begin{figure}
	\centering
	\begin{tikzpicture}
	\node[anchor=south west,inner sep=0] at (0,0) {
		\includegraphics[width=0.495\textwidth, clip=true, trim={2cm 0cm 1cm 0cm}]{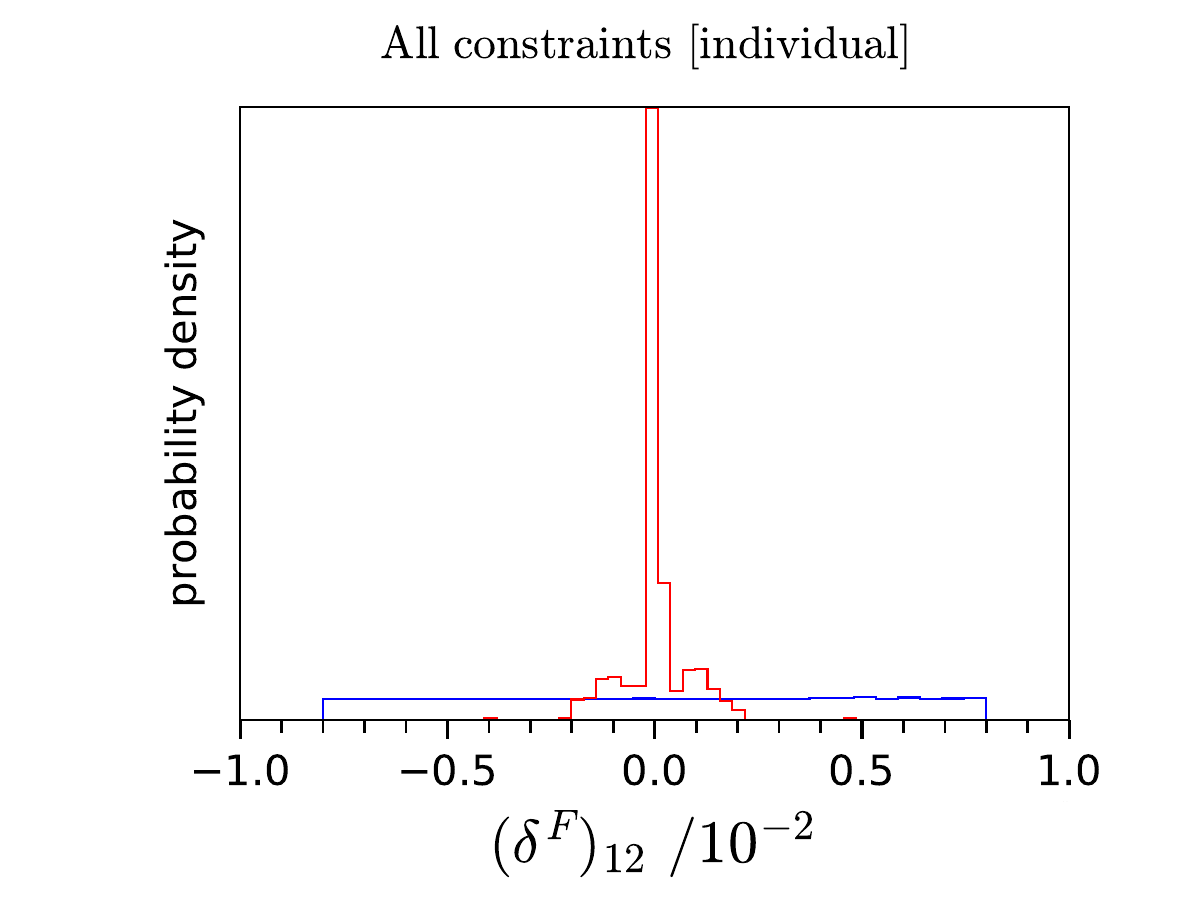} 
		\includegraphics[width=0.495\textwidth, clip=true, trim={2cm 0cm 1cm 0cm}]{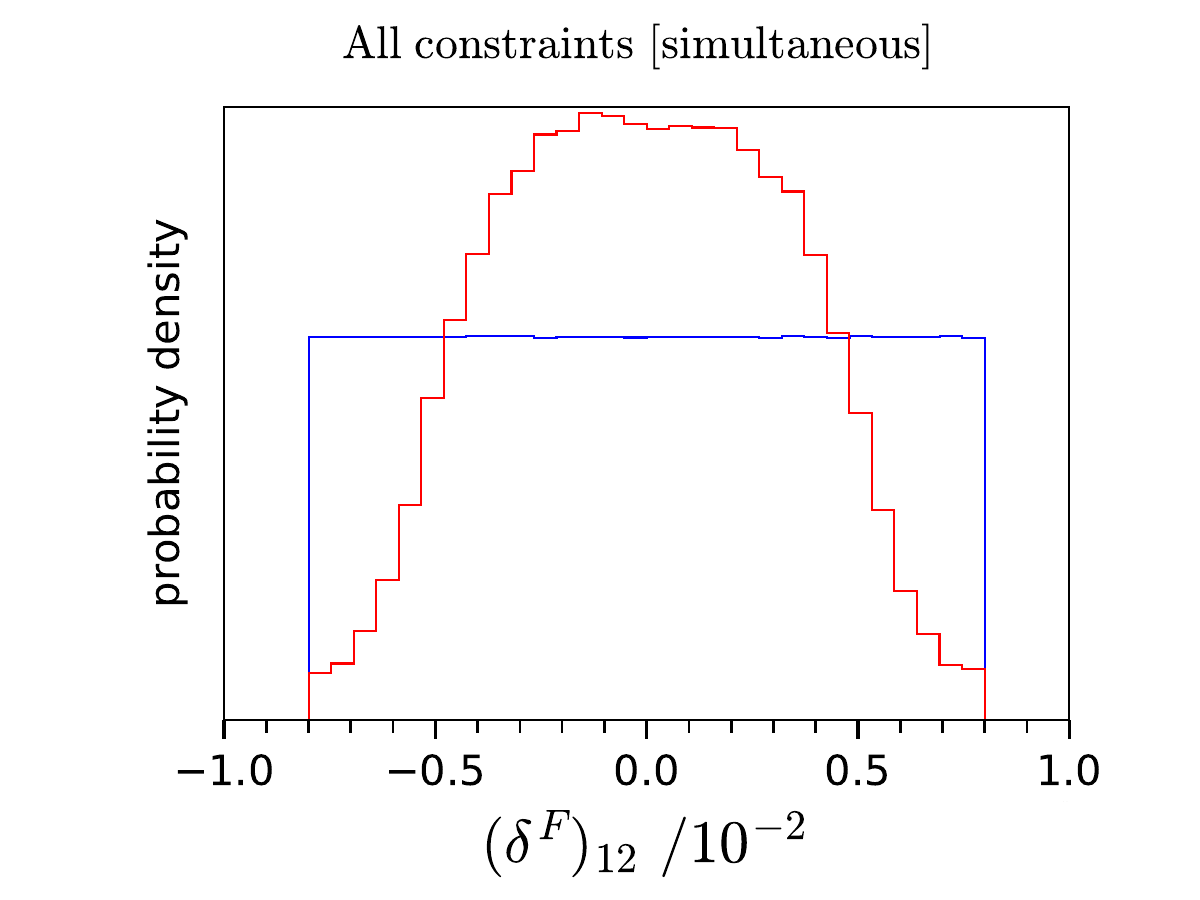}
	};
	\node at (1.3,5.5) {\textbf{a)}};
	\node at (8.8,5.5) {\textbf{b)}};
	\end{tikzpicture}
	\caption{Comparison of individual (panel a)) vs.\ simultaneous (panel b)) scan of the NMFV parameter $(\deltaf)_{12}$ around Scenario 1. Each panel shows the prior (blue) together with the posterior (red) distributions.}
	\label{fig:comparison_plots_deltaf_12}
	\vspace*{\floatsep}	
	\centering
	\begin{tikzpicture}
		\node[anchor=south west,inner sep=0] at (0,0) {
			\includegraphics[width=0.495\textwidth, clip=true, trim={2cm 0cm 1cm 0cm}]{Simultaneous_BP1_delta_F12_AllCst.pdf}
			\includegraphics[width=0.495\textwidth, clip=true, trim={2cm 0cm 1cm 0cm}]{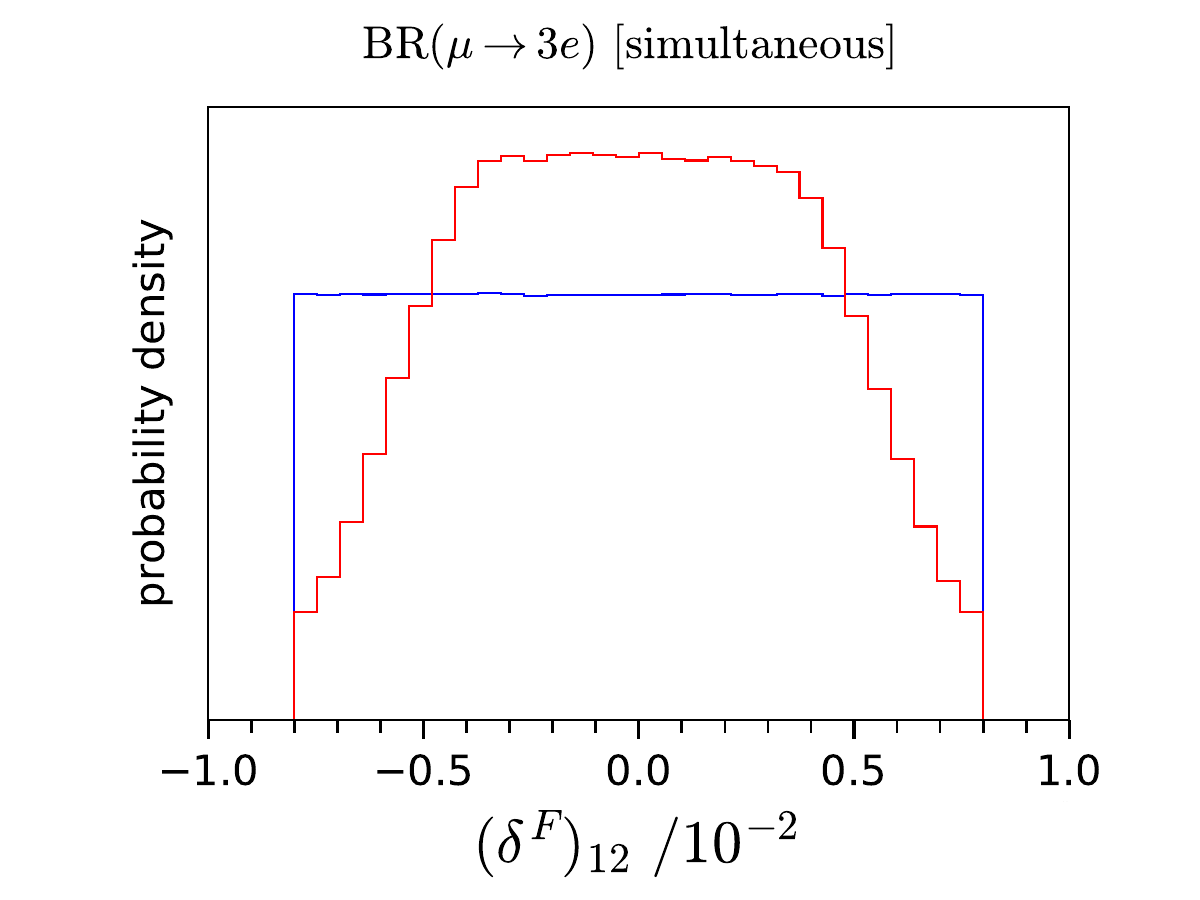}
		};
		\node at (1.2,5.5) {\textbf{a)}};
		\node at (8.7,5.5) {\textbf{b)}};
	\end{tikzpicture}
	\caption{Dominant constraints on the parameter $(\deltaf)_{12}$ from simultaneous scan around Scenario 1. Prior distributions are given in blue and posterior distribution are given in red.}
	\label{fig:BP1_deltaf_12}
\end{figure}

We discuss here in detail the results obtained for the full NMFV scan around the reference scenario 1. The MFV parameters are fixed at the values given in Table \ref{tab:RefScenarioGUT}, while we scan over the NMFV parameters according to the ranges given in Table \ref{Tab:ParameterRange}, either individually (i.e.\ keeping all but one parameter to zero), or simultaneously. For each performed scan, we record the prior distribution containing all points featuring a physical mass spectrum and neutralino dark matter candidate (see also Fig.\ \ref{Fig:Flowchart}) as well as the posterior distribution obtained when imposing either one or all constraints summarized in Table \ref{Tab:Constraints}.

Fig.\ \ref{fig:comparison_plots_deltaf_12} shows the obtained prior and posterior distributions for the NMFV parameter $(\deltaf)_{12}$. The viable region for this parameter with respect to the imposed constraints is much larger for the case of the simultaneous scan as compared to the individual scan result. Indeed, it is possible that more than one of the NMFV parameters enters the calculation of one or more observables. In such a case, interferences and/or cancellations between the contributions induced by different NMFV parameters can occur. As a consequence, they give rise to viable regions of parameter space that would not be fully explored when varying each parameter in isolation. This is seen quantitatively as a broadening of posterior distributions when comparing a simultaneous scan result against a histogram from an individual scan. Let us emphasize that this feature is present for several of the flavour violating parameters under consideration in our study.

Fig.\ \ref{fig:BP1_deltaf_12}, panel b) shows the action of a single observable, ${\rm BR}(\mu \to 3e)$, on the same parameter $(\deltaf)_{12}$ for simulultaneous scan, and can thus be directly compared to Fig.\ \ref{fig:comparison_plots_deltaf_12} \footnote{Note that panel b) of Fig.\ \ref{fig:comparison_plots_deltaf_12} is identical to panel a) in Fig.\ \ref{fig:BP1_deltaf_12}.}. Since the shape of the single-constraint posterior almost matches the posterior obtained imposing all constraints, we conclude that this parameter is mainly limited by the $\mu\rightarrow 3e$ lepton decay bound. The $\mu \to e\gamma$ observable is less important in this case (see Table \ref{Tab:constrained_parameters_range}, corresponding posterior not shown).

\begin{figure}[H]
	\centering
	\begin{tikzpicture}
		\node[anchor=south west,inner sep=0] at (0,0) {
			\includegraphics[width=0.495\textwidth, clip=true, trim={2cm 0cm 1cm 0cm}]{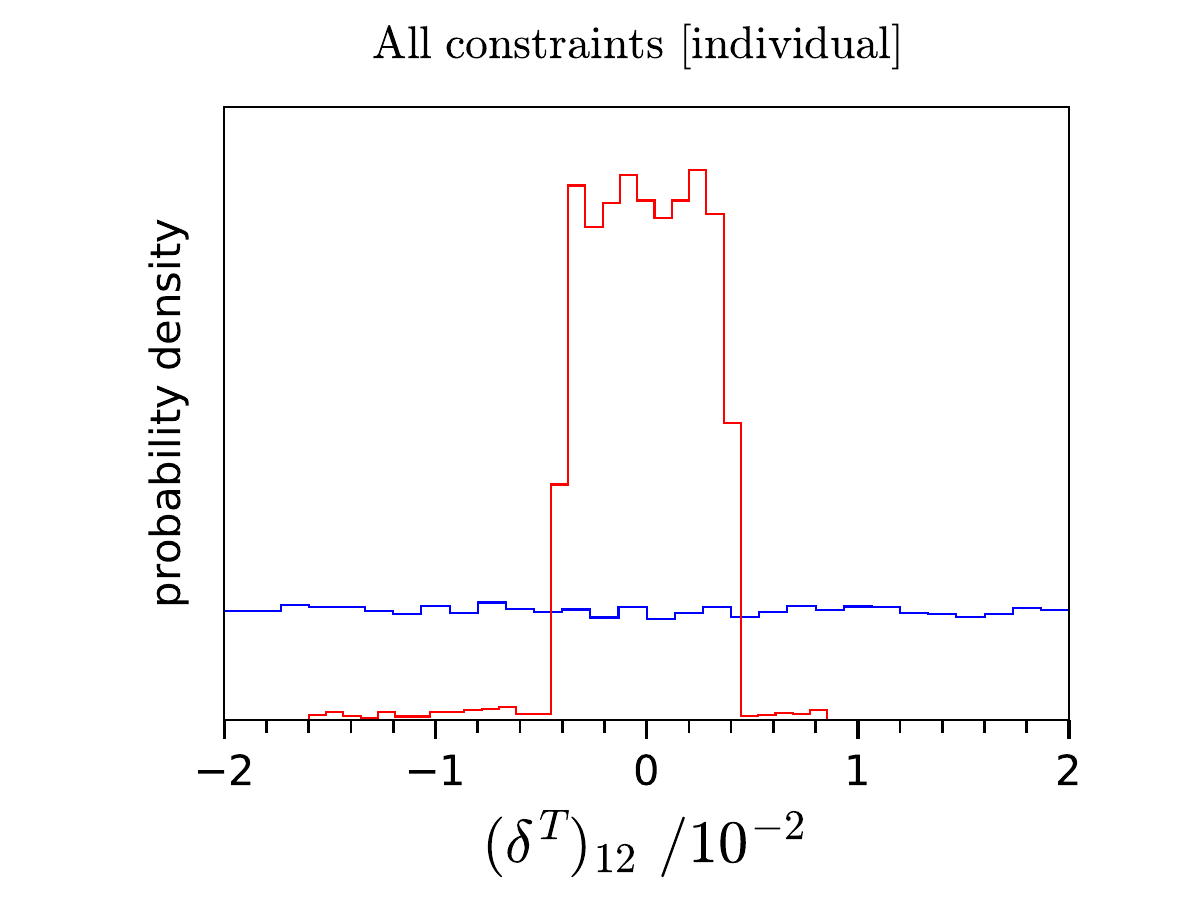} 
			\includegraphics[width=0.495\textwidth, clip=true, trim={2cm 0cm 1cm 0cm}]{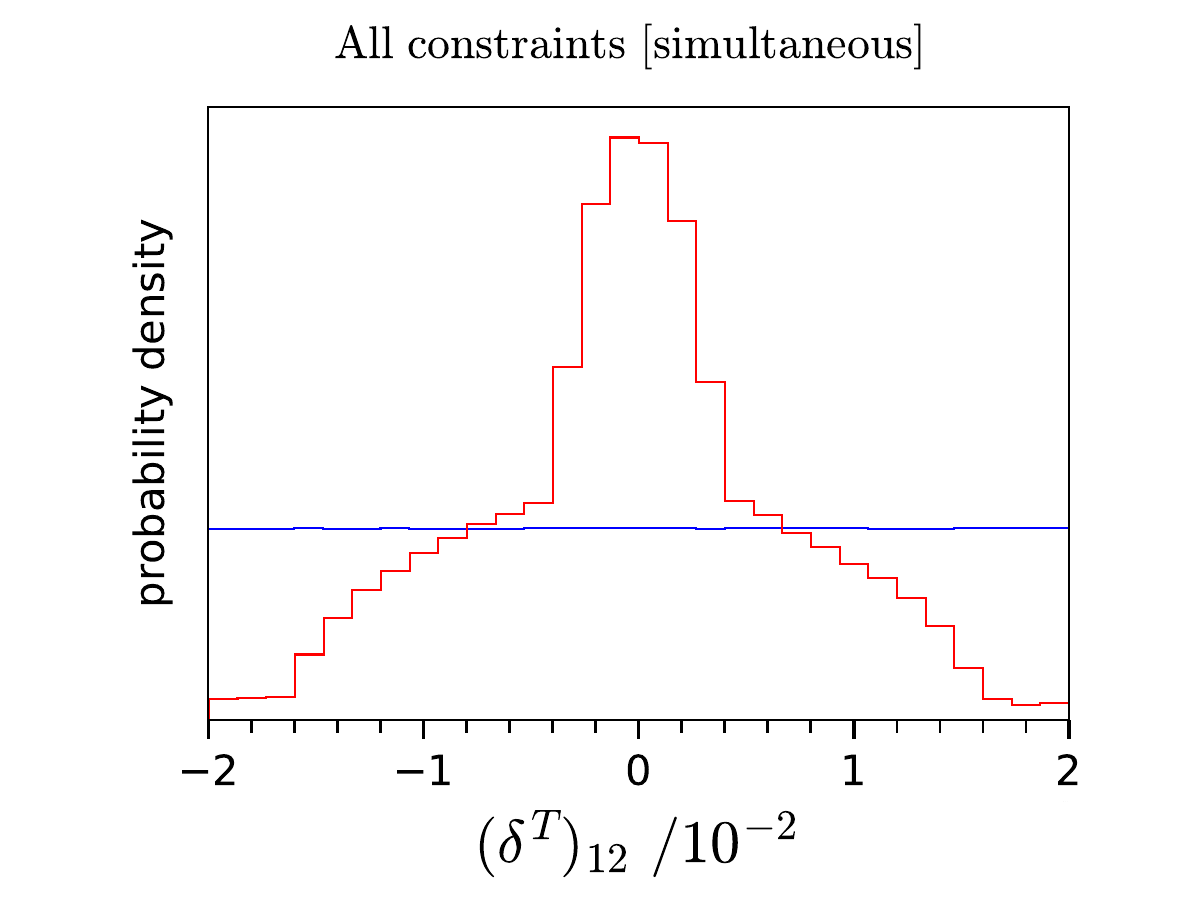}
		};
		\node at (1.2,5.5) {\textbf{a)}};
		\node at (8.7,5.5) {\textbf{b)}};
	\end{tikzpicture}
	\caption{Comparison of individual (left) vs.\ simultaneous (right) scan of the NMFV parameter $(\deltat)_{12}$ around Scenario 1. Each panel shows the prior (blue) together with the posterior (red) distributions.} 
	\label{fig:comparison_plots_deltat_12}
	\vspace*{\floatsep}
	\centering
	\begin{tikzpicture}
		\node[anchor=south west,inner sep=0] at (0,0) {
			\includegraphics[width=0.495\textwidth, clip=true, trim={2cm 0cm 1cm 0cm}]{Simultaneous_BP1_delta_T12_AllCst.pdf}
			\includegraphics[width=0.495\textwidth, clip=true, trim={2cm 0cm 1cm 0cm}]{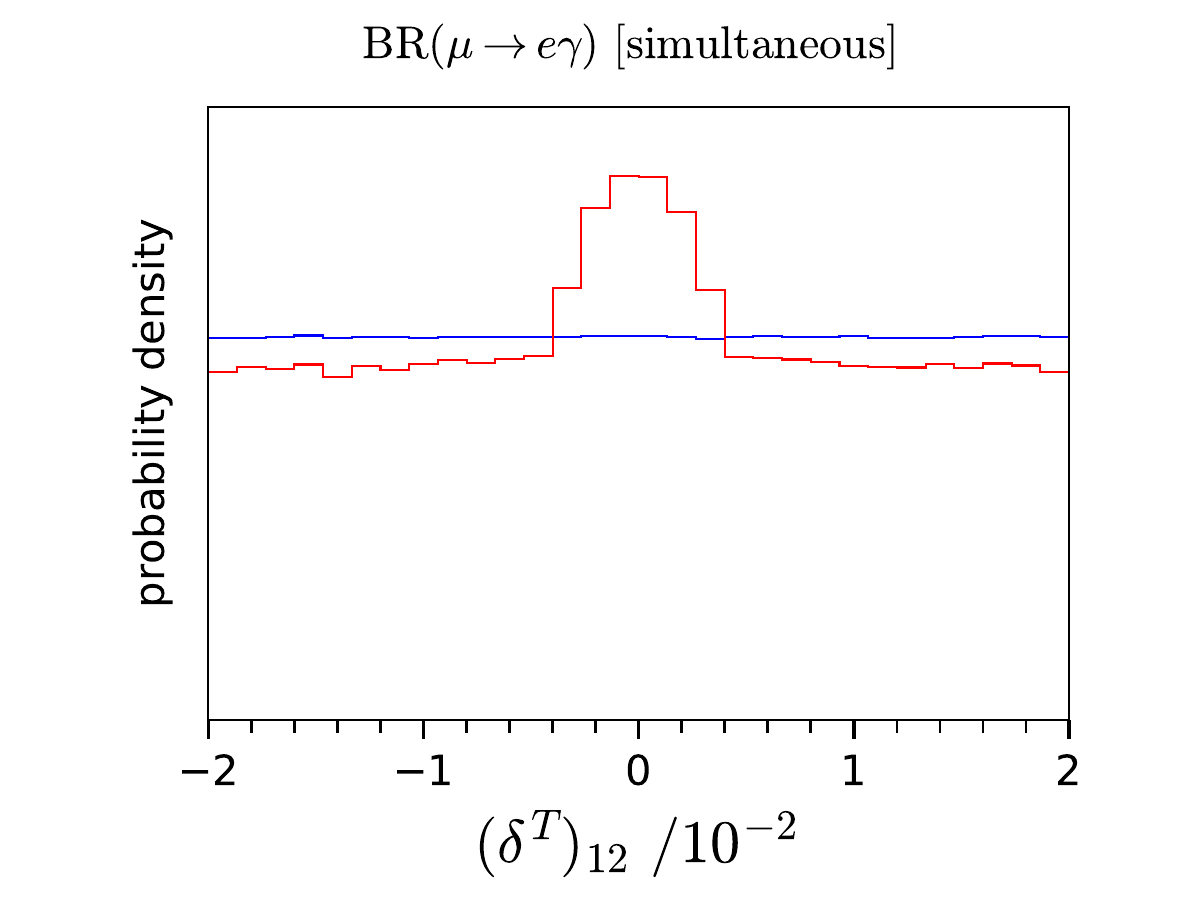}};
		\node[anchor=south west,inner sep=0] at (0,-6.8) {	
			\includegraphics[width=0.495\textwidth, clip=true, trim={2cm 0cm 1cm 0cm}]{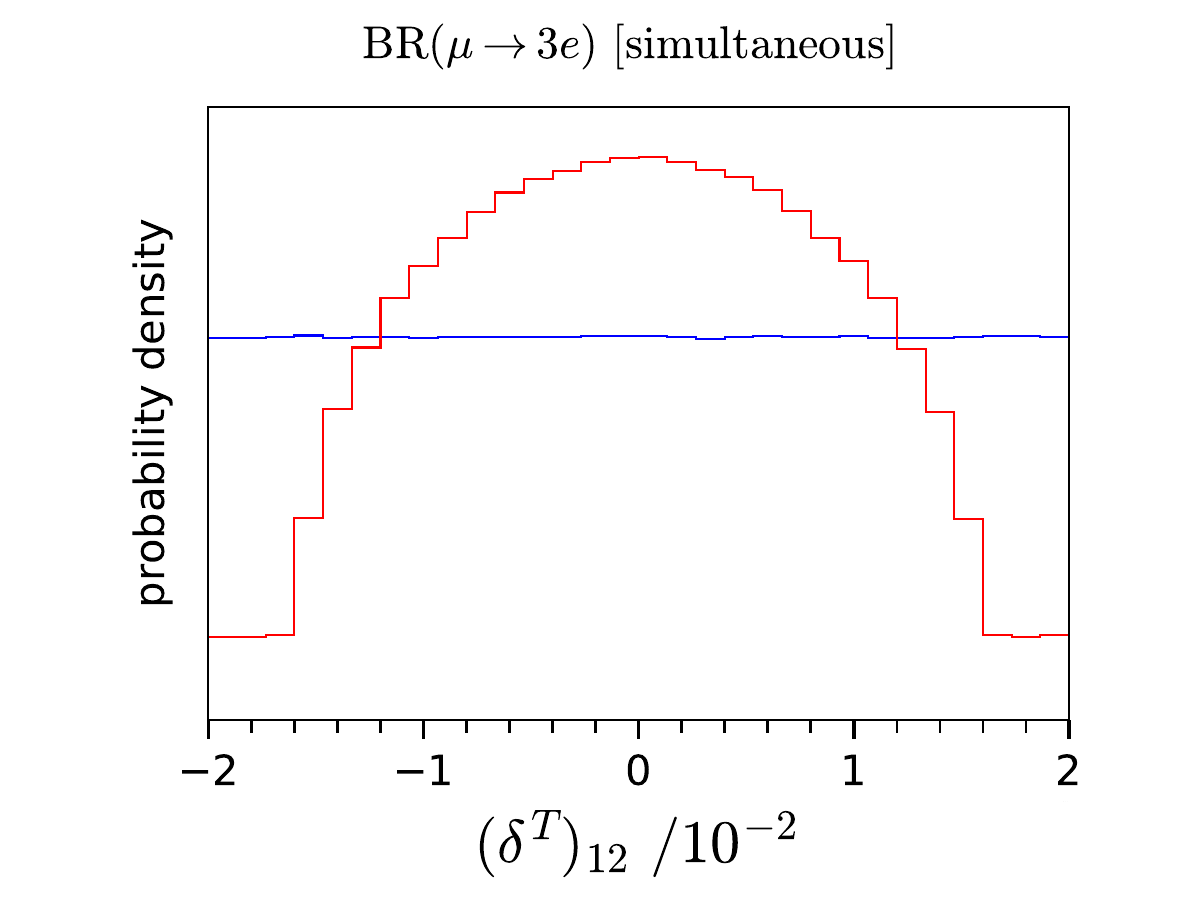}
			\includegraphics[width=0.495\textwidth, clip=true, trim={2cm 0cm 1cm 0cm}]{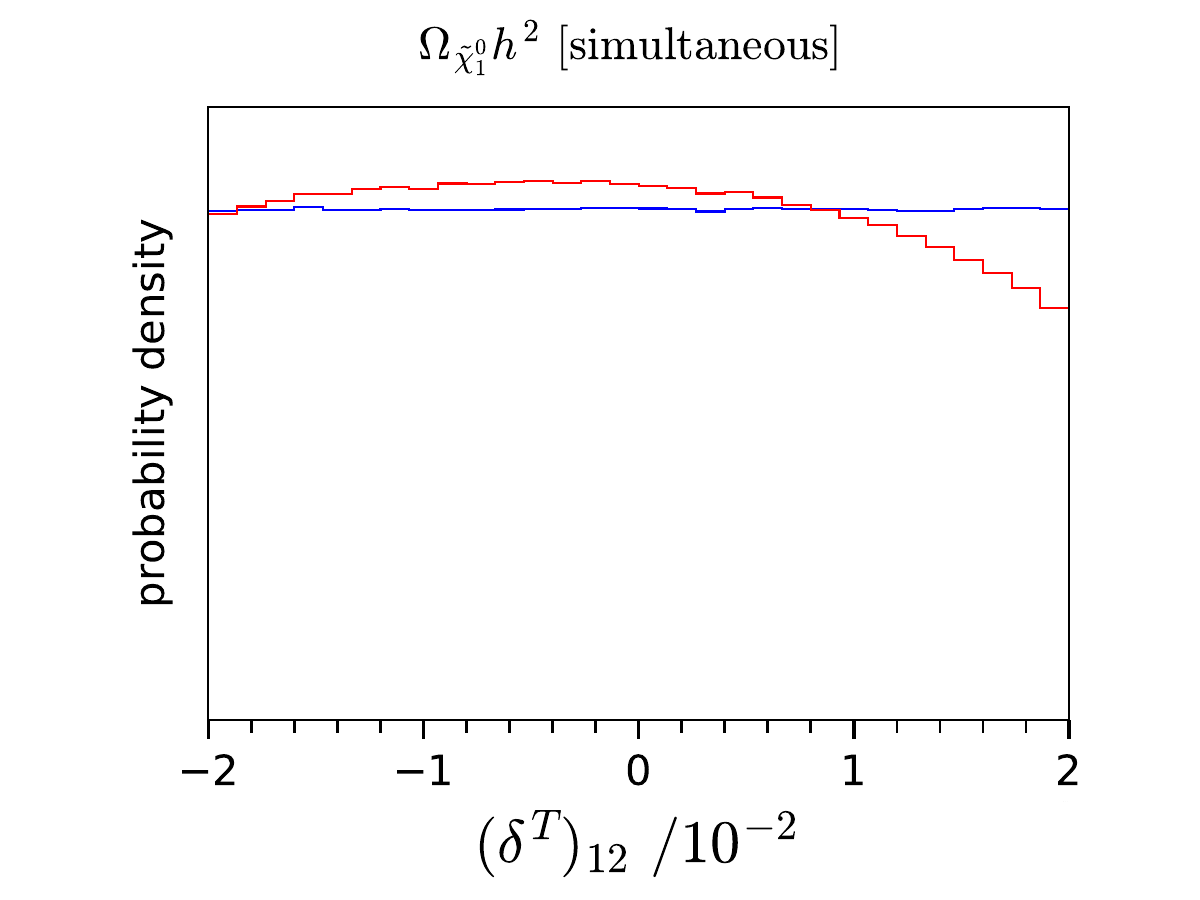}
		};
		\node at (1.1,5.5) {\textbf{a)}};
		\node at (8.7,5.5) {\textbf{b)}};
		\node at (1.1,-1.3) {\textbf{c)}};
		\node at (8.7,-1.3) {\textbf{d)}};
	\end{tikzpicture}
	\caption{Dominant constraints on the parameter $(\deltat)_{12}$ from simultaneous scan around Scenario 1. Prior distributions are given in blue and posterior distribution are given in red.}
	\label{fig:BP1_deltat_12}
\end{figure}

Coming to the parameter $(\deltat)_{12}$ shown in Fig.\ \ref{fig:comparison_plots_deltat_12} including all constraints, note that the obtained viable interval is again broadened when comparing the individual scan, leading to $|(\deltat)_{12}| \lesssim 0.2\times10^{-2}$, with the simultaneous one yielding the range $|(\deltat)_{12}| \lesssim 1.6\times10^{-2}$. For the same NMFV parameter $(\deltat)_{12}$, we detail in Fig.\ \ref{fig:BP1_deltat_12} the effect of the three most important experimental constraints in the simultaneous scan. The $\mu\rightarrow e\gamma$ constraint can be seen to admit the entirety of the scanned region of parameter space in the simultaneous scan, whereas it is far the most stringent constraint in the individual scan (see Fig.\ \ref{fig:comparison_plots_deltat_12}). Indeed, $\mu\rightarrow 3e$ is the most constraining observable for this parameter when varied along with other flavour violating entries of mass matrices. In addition, Fig.\ \ref{fig:BP1_deltat_12} illustrates how the obtained shape of the posterior distribution is due to the influence of three experimental constraints imposed on the parameter space.

\begin{figure}
	\centering
	\begin{tikzpicture}
		\node[anchor=south west,inner sep=0] at (0,0) {		
			\includegraphics[width=0.495\textwidth, clip=true, trim={2cm 0cm 1cm 0cm}]{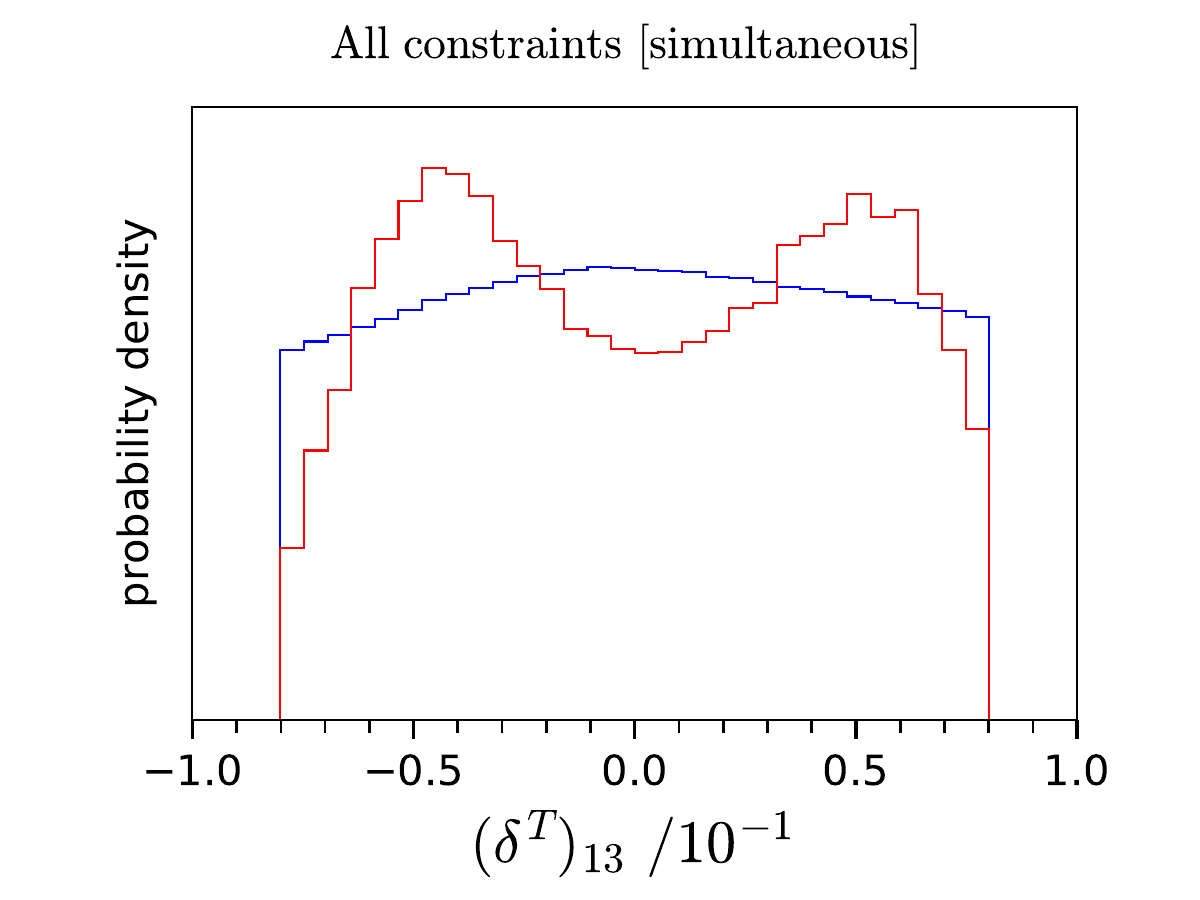}
			\includegraphics[width=0.495\textwidth, clip=true, trim={2cm 0cm 1cm 0cm}]{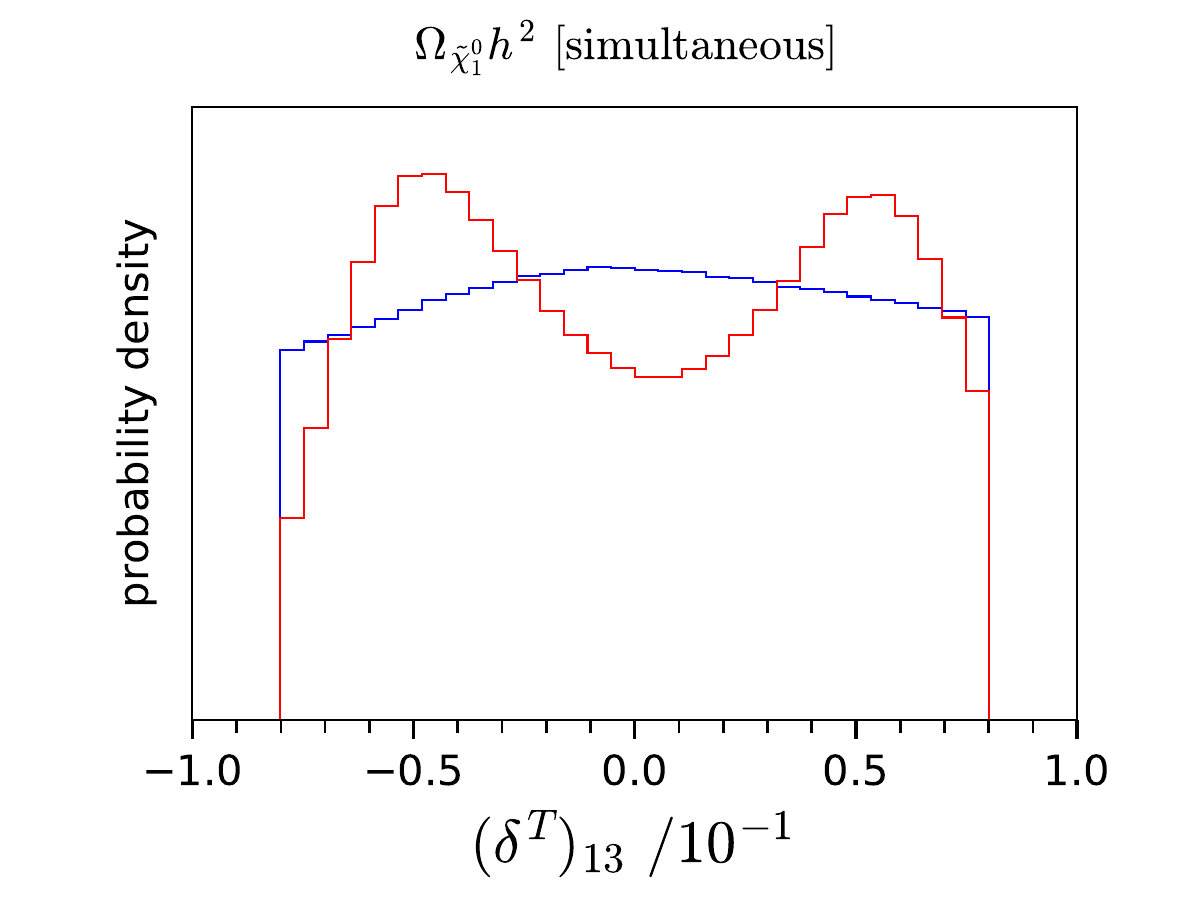}
		};
		\node at (1.1,5.5) {\textbf{a)}};
		\node at (8.7,5.5) {\textbf{b)}};
	\end{tikzpicture}
	\caption{Dominant constraints on the parameter $(\deltat)_{13}$ from simultaneous scan around Scenario 1. Prior distributions are given in blue and posterior distribution are given in red.}
	\label{fig:BP1_deltat_13}
\end{figure}

We now discuss the parameter $(\deltat)_{13}$ shown in Fig.\ \ref{fig:BP1_deltat_13}. We can notice that it is constrained only by the neutralino relic density and that the flavour constraints have no effect. This gives insight on the unexpected shape of the posterior distribution: As we have seen for two examples above, other NMFV parameters are allowed under flavour constraints to shift significantly away from zero. This has a marked effect in reducing superpartner masses which are determined by diagonalising the mass-squared matrices from Eq.\ \eqref{Eq:Full_MSSM_soft_matrices}. This applies in particular to the ``right-handed'' smuon mass, as the initial smallness of $m_{T_2}$ means that small NMFV parameters can slightly lower the smuon mass. As a further consequence, the relic density is then reduced due to the smaller mass difference between smuon and neutralino, which increases the importance of co-annihilation and smuon pair annihilation. However, the smuon mass also is influenced by $(\deltat)_{13}$, which by virtue of being unconstrained by flavour observables, may be non-zero. Moreover, this particular parameter increases the lightest smuon mass due to the specific hierarchies in the mass matrix. The smuon mass being decreased by other non-zero NMFV parameters, $(\deltat)_{13}$ being non-zero then re-establishes the initial mass difference between the smuon and neutralino allowing the relic density to stay within the Planck limits. If one relaxes the assumption that the neutralino $\tilde{\chi}_1^0$ is the only dark matter candidate, i.e.\ relax the lower limit on the relic density, then the caracteristic shape observed for $(\deltat)_{13}$ in Fig.\ \ref{fig:BP1_deltat_13} disappears. 

Any NMFV parameters among those listed in Table \ref{Tab:ParameterRange} whose distributions are not detailed here do not have any interesting phenomenona associated with the imposed constraints, therefore the reader can deduce the full effect and resulting ranges from Table \ref{Tab:constrained_parameters_range}. Recall that for this scenario, the parameters $(\deltat)_{13}$, $(\deltaft)_{21}$, and $(\deltaft)_{32}$ have been set to zero due to requirements for a physical spectrum and neutral LSP.
For all $\deltatt$ parameters, the main requirements are for a physically relevant spectrum and uncharged LSP, hence we conclude that the \textit{prior} distribution dominantes over flavour observables that we test against here. Finally, we do not discuss the $\deltaft$ parameters as the corresponding results are much the same as for the scan around Scenario 2 presented in the following. 

From the discussed results related to reference Scenario 1, it is clear that varying the NMFV parameters individually is not sufficient to properly explore the entirety of parameter space. For this reason, we do not discuss individual variations any further.

\subsection{Scan around Scenario 2}

\begin{figure}
	\centering
	\begin{tikzpicture}
		\node[anchor=south west,inner sep=0] at (0,0) {				
			\includegraphics[width=0.495\textwidth, clip=true, trim={2cm 0cm 1cm 0cm}]{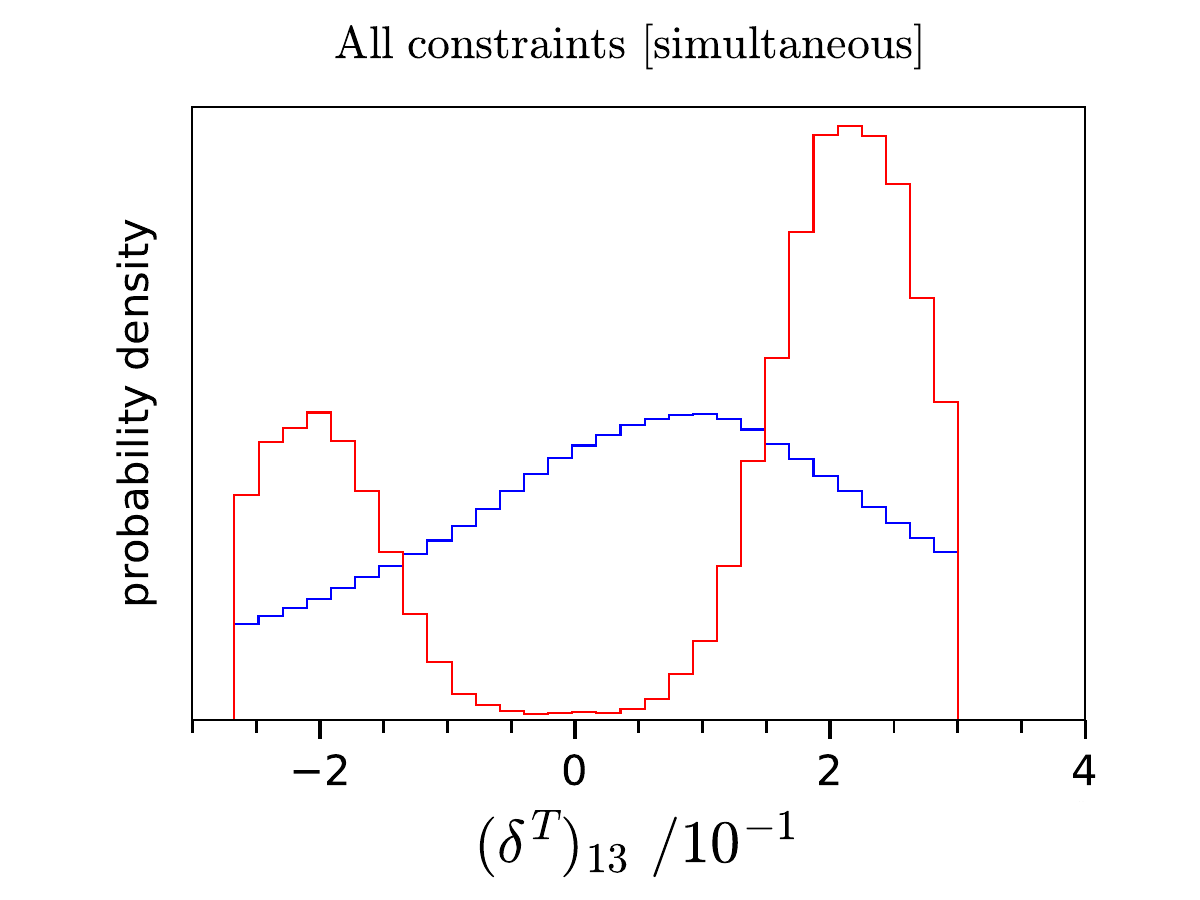} 
			\includegraphics[width=0.495\textwidth, clip=true, trim={2cm 0cm 1cm 0cm}]{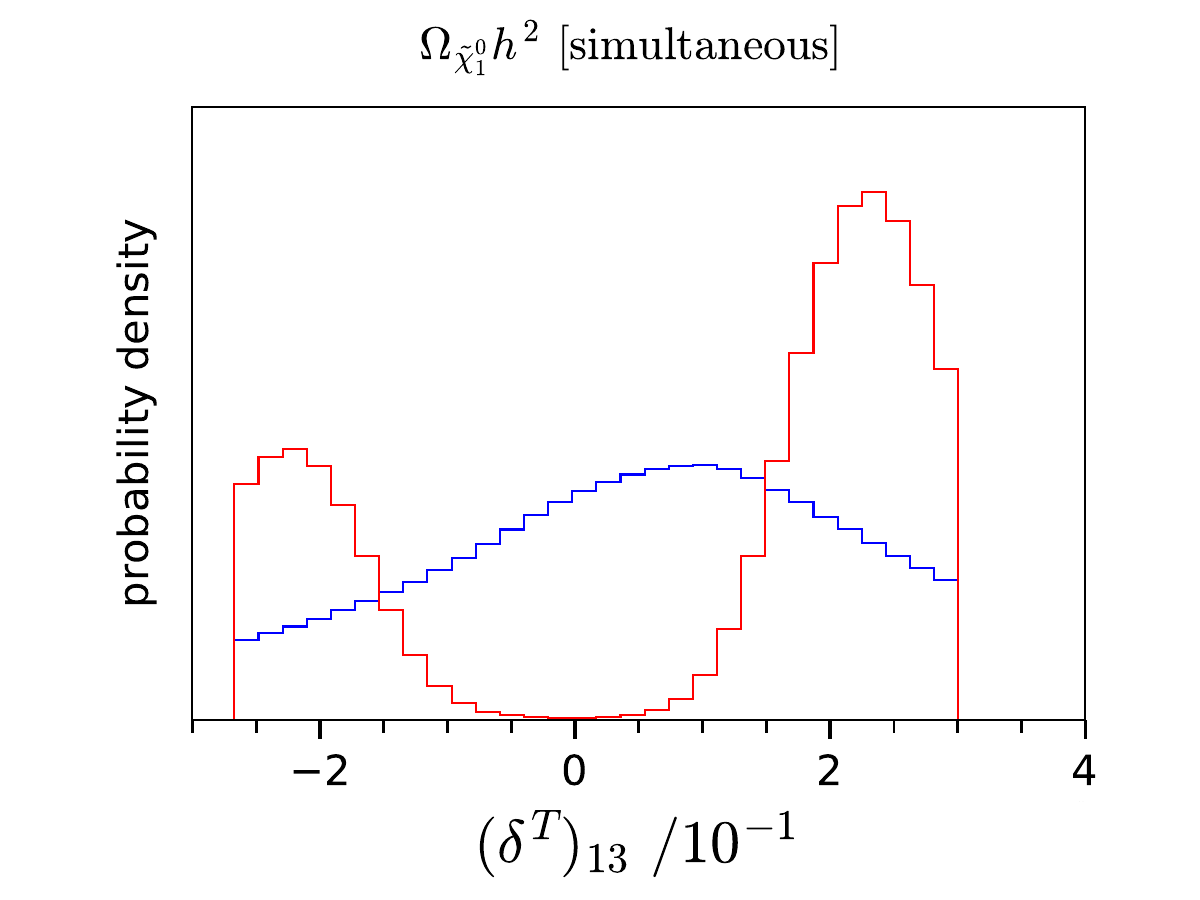} 
		};
		\node at (1.1,5.5) {\textbf{a)}};
		\node at (8.7,5.5) {\textbf{b)}};
	\end{tikzpicture}
	\caption{Dominant constraints on the parameter $(\deltat)_{13}$ from simultaneous scan around Scenario 2. Prior distributions are given in blue and posterior distribution are given in red.}
	\label{fig:BP2_deltat_13}
\end{figure}

Here, we discuss selected results of the simultaneous scan of all 15 NMFV parameters around Scenario 2. NMFV parameters are varied according to the ranges given in Table \ref{Tab:ParameterRange}, while the MFV parameters are fixed to the values given in Table \ref{tab:RefScenarioGUT}. Note that the change of the MFV parameters as compared to Scenario 1 allows the variation of all 15 NMFV parameters, while three of them were set to zero for Scenario 1. This yields limits on the full range of flavour violation allowed in Scenario 2. 

\begin{figure}[H]
	\centering
	\begin{tikzpicture}
		\node[anchor=south west,inner sep=0] at (0,0) {			
			\includegraphics[width=0.495\textwidth, clip=true, trim={2cm 0cm 1cm 0cm}]{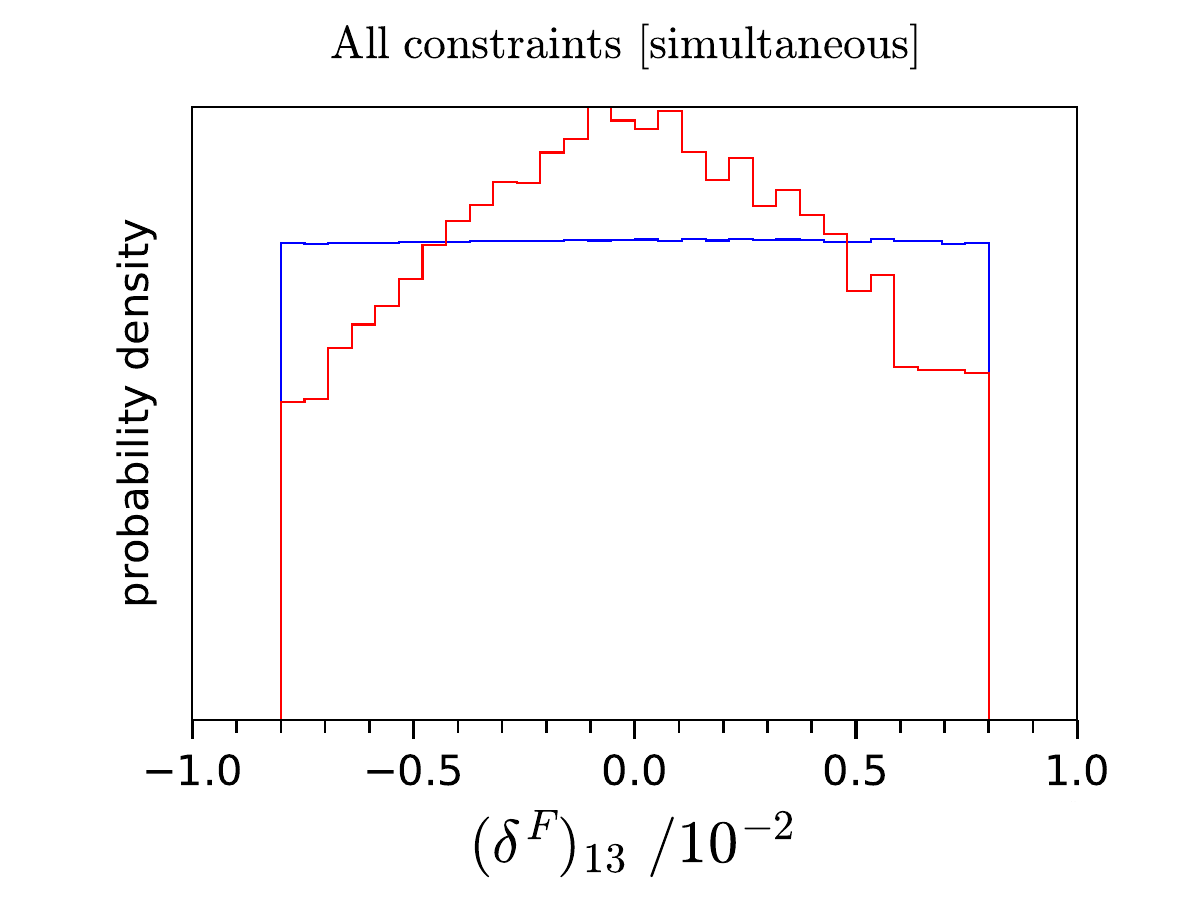} 
			\includegraphics[width=0.495\textwidth, clip=true, trim={2cm 0cm 1cm 0cm}]{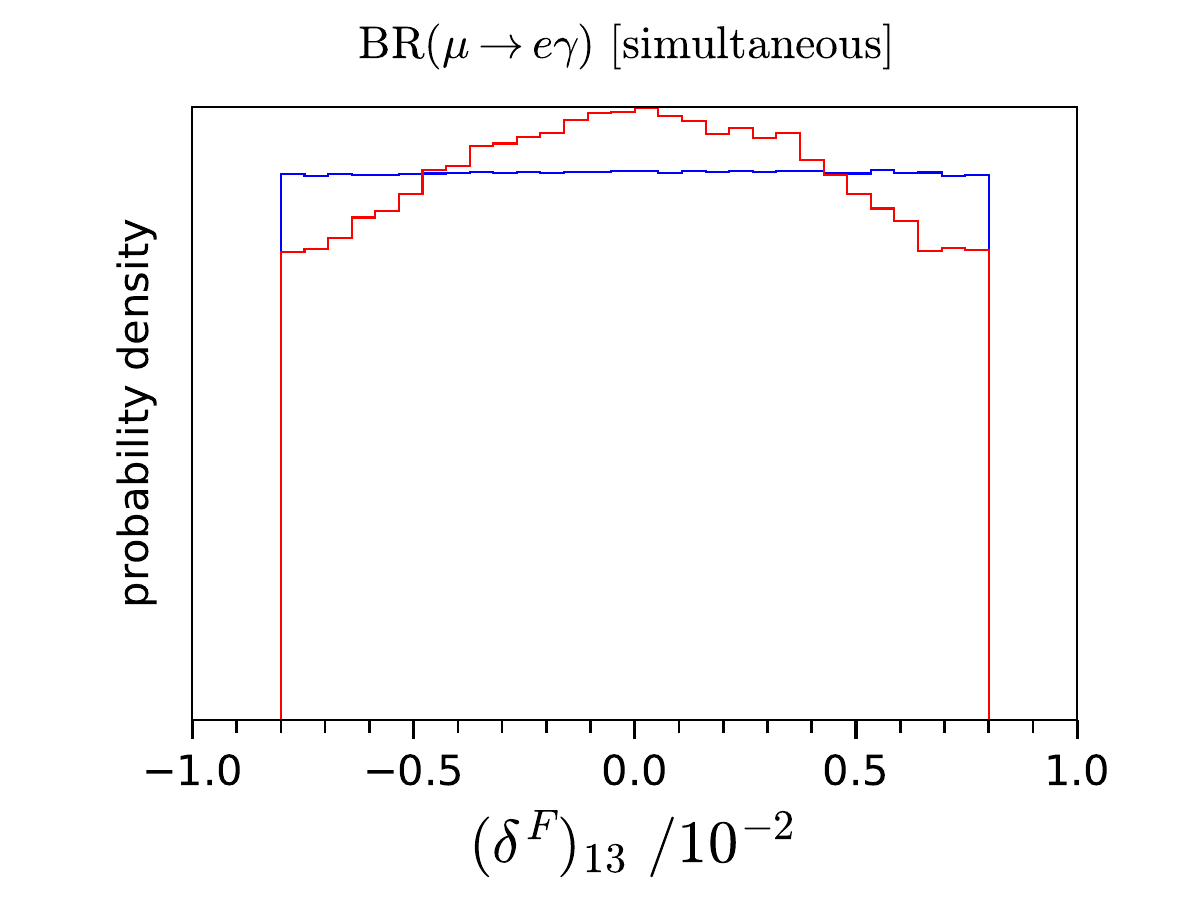} 
		};
		\node[anchor=south west,inner sep=0] at (0,-6.8) {	
		\includegraphics[width=0.495\textwidth, clip=true, trim={2cm 0cm 1cm 0cm}]{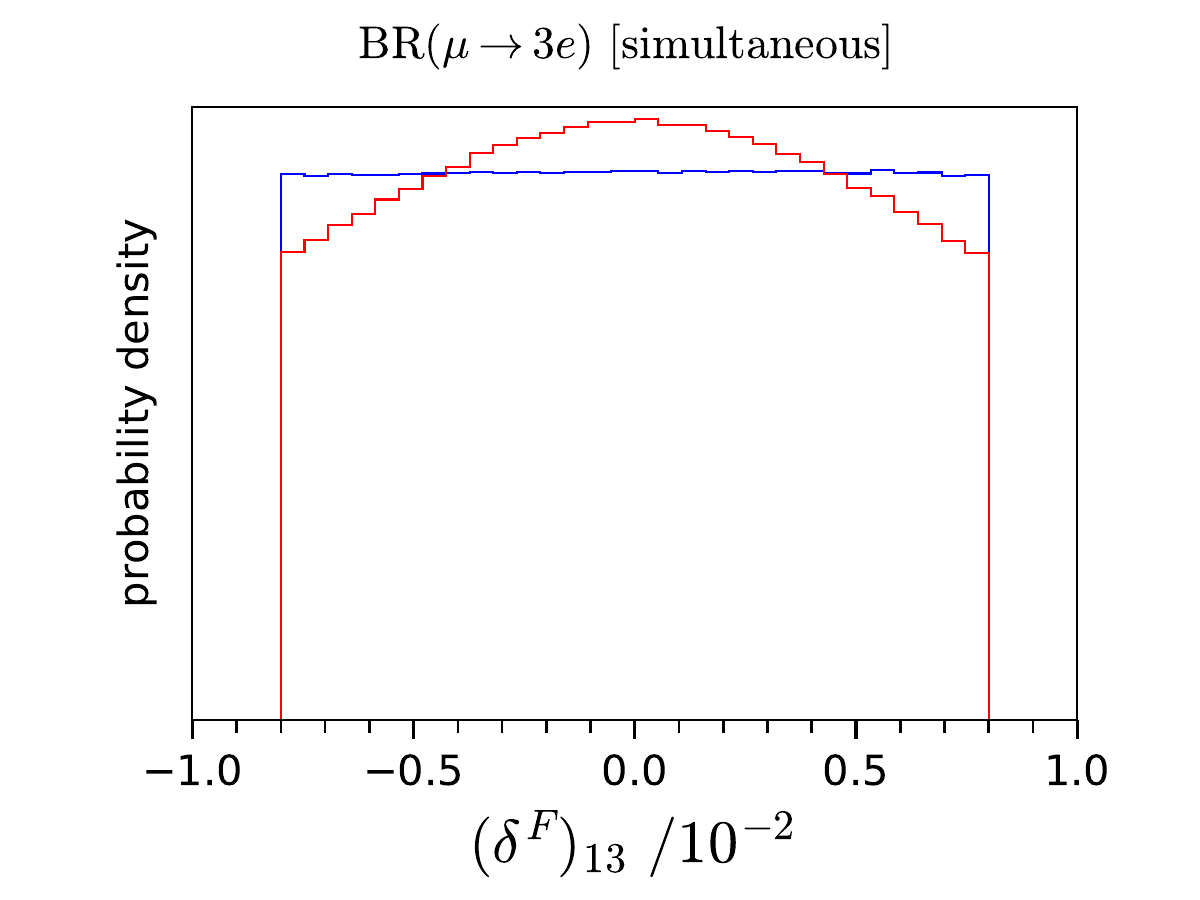}
		\includegraphics[width=0.495\textwidth, clip=true, trim={2cm 0cm 1cm 0cm}]{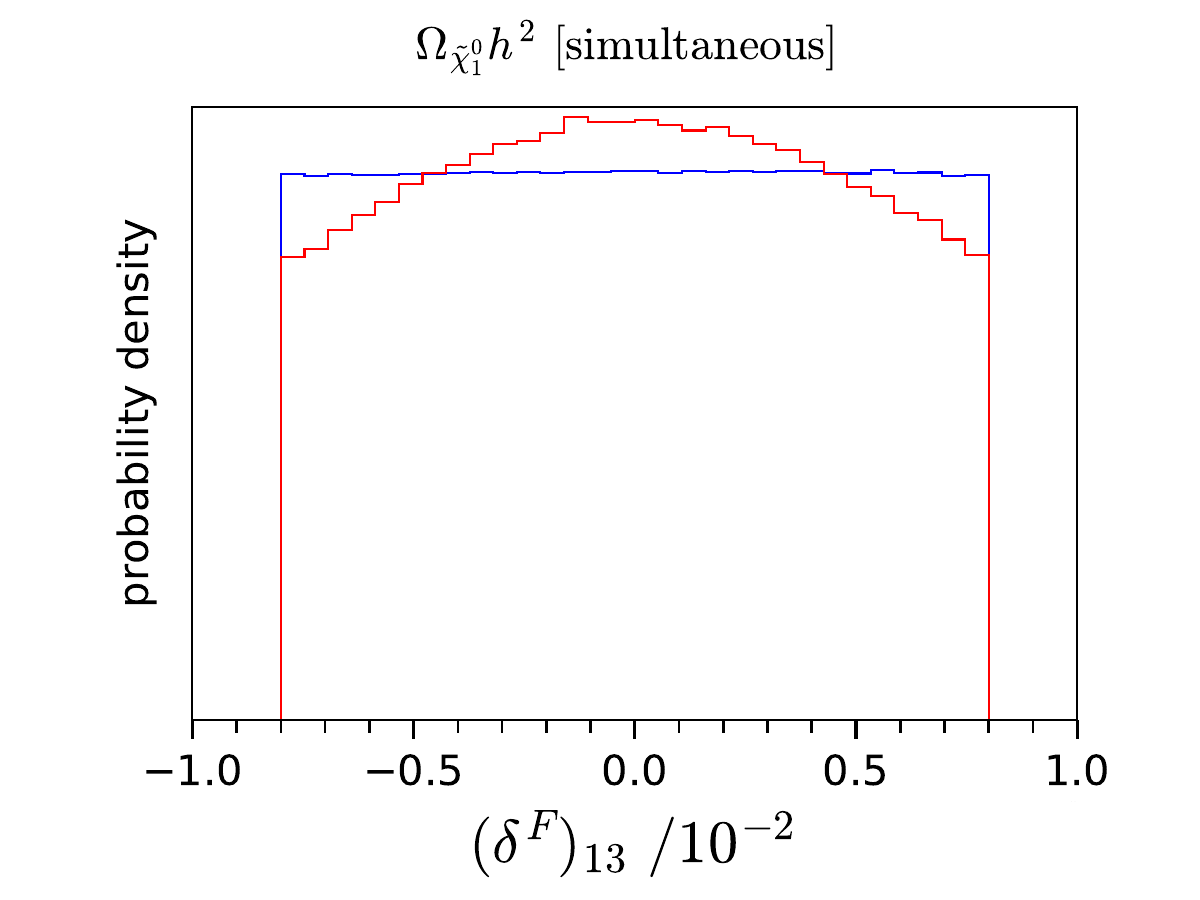}};
		\node at (1.0,5.5) {\textbf{a)}};
		\node at (8.6,5.5) {\textbf{b)}};
		\node at (1.0,-1.3) {\textbf{c)}};
		\node at (8.6,-1.3) {\textbf{d)}};
	\end{tikzpicture}
	\caption{Dominant constraints on the parameter $(\deltaf)_{13}$ from simultaneous scan around Scenario 2. Prior distributions are given in blue and posterior distribution are given in red.}
	\label{fig:BP2_deltaf_13}
	\vspace*{\floatsep}
	\centering
	\begin{tikzpicture}
		\node[anchor=south west,inner sep=0] at (0,0) {						
			\includegraphics[width=0.495\textwidth, clip=true, trim={2cm 0cm 1cm 0cm}]{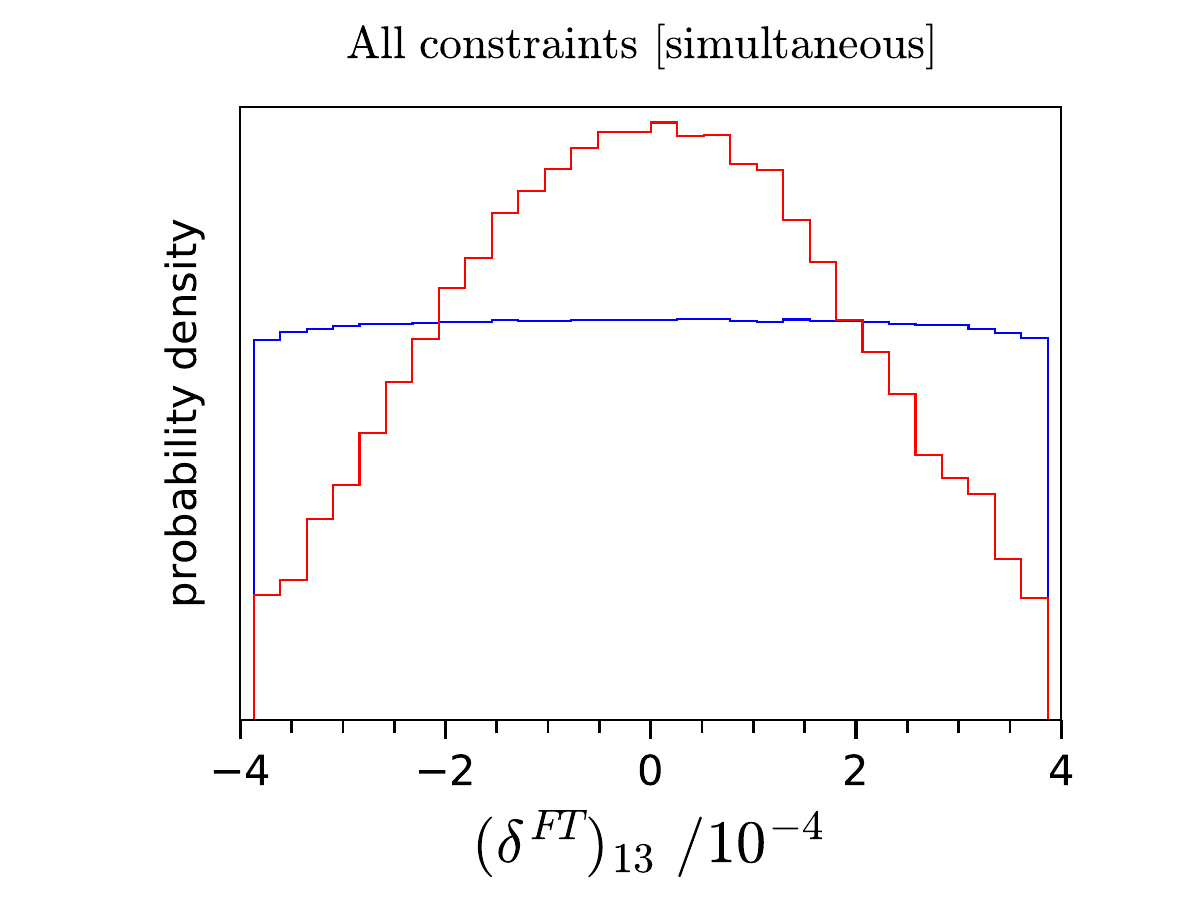} 
			\includegraphics[width=0.495\textwidth, clip=true, trim={2cm 0cm 1cm 0cm}]{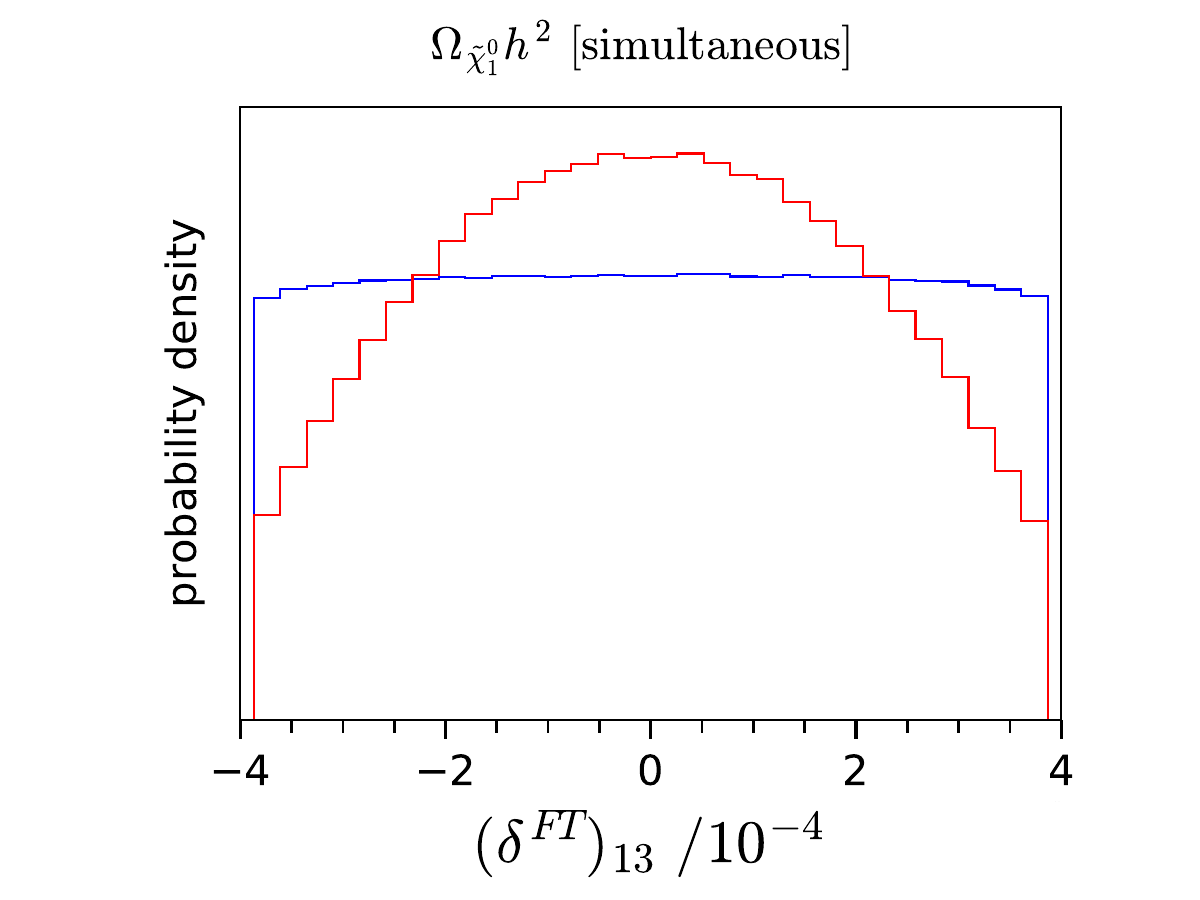}
		};
		\node at (1.25,5.5) {\textbf{a)}};
		\node at (8.85,5.5) {\textbf{b)}};
	\end{tikzpicture}
	\caption{Dominant constraints on the parameter $(\deltaft)_{13}$ from simultaneous scan around Scenario 2. Prior distributions are given in blue and posterior distribution are given in red.}
	\label{fig:BP2_deltaft_32}
\end{figure}

Starting the discussion with the parameter $(\deltat)_{13}$ for which we present the resulting prior and posterior distributions in Fig.\ \ref{fig:BP2_deltat_13}, we observe the same feature as for Scenario 1 (see Fig.\ \ref{fig:BP1_deltat_13}), but more pronounced. Again, slightly positive or negative values for $(\deltat)_{23}$ counteract the effects of other NMFV parameters on the neutralino relic density as explained previously. 

Coming to the parameter $(\deltaf)_{13}$, Fig.\ \ref{fig:BP2_deltaf_13} shows that, rather than a single observable having a clear effect, cumulatively $\mu\rightarrow e\gamma$, $\mu\rightarrow 3e$, and $\Omega_{\tilde{\chi}^0_1}h^2$ constrain the parameter together with each having a similar effect. Here, we see particularly the effect of flavour violating muon decays on $(\delta)_{13}$ parameters as elaborated upon in the beginning of Section \ref{Sec:Results}.

In the same way as for Scenario 1, all $\deltatt$ parameters are constrained by the ``prior'' requirement of a physical mass spectrum and a neutralino dark matter candidate. Flavour observables have a negligible effect (see Table \ref{Tab:constrained_parameters_range}).

An example of the posterior distribution for $\deltaft$ parameters is shown in Fig.\ \ref{fig:BP2_deltaft_32}, namely for $(\deltaft)_{13}$. This parameter is constrained almost entirely by the relic density bound, as can be seen in the similarity of the two panels. Let us recall that complete information on limits and dominant constraints of all NMFV parameters associated with Scenario 2 is summarized in Table \ref{Tab:constrained_parameters_range}.

\subsection{SUSY Scale NMFV parameters for Scenario 2}

While from the model-building point of view it is useful to explore the allowed level of flavour violation at the GUT scale, it is equally important to explore the resulting physics at the SUSY scale. Renormalization group running from the GUT scale to the SUSY scale will break the unification conditions given in Eq.\ \eqref{Eq:matrix_defintions} and consequently in Eq.\ \eqref{Eq:NMFV_GUT_A4xSU5_deltas}. The fact that these relations are not valid any more below the GUT scale is an essential and intrinsic part of Grand Unification. The present Section is devoted to highlighting selected results related to the NMFV parameters obtained at the SUSY scale. More precisely, we study the behaviour of different SUSY scale NMFV parameters which stem from a single NMFV parameter at the GUT scale.  

\begin{figure}[H]
	\centering
	\begin{tikzpicture}
		\node[anchor=south west,inner sep=0] at (3.7,0) {	\includegraphics[width=0.495\textwidth, clip=true, trim={2cm 0cm 1cm 0cm}]{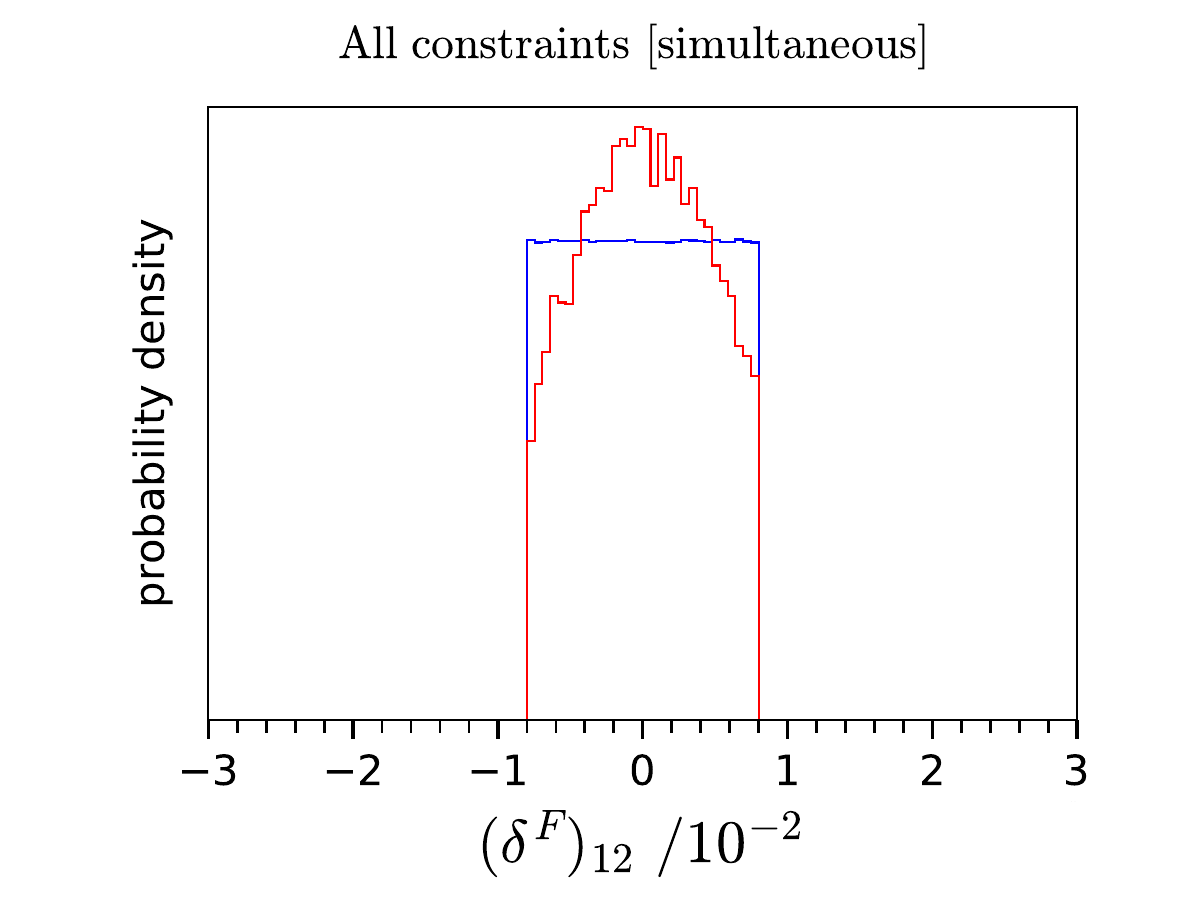}};
		\node[anchor=south west,inner sep=0] at (0,-6.8) {	
		\includegraphics[width=0.495\textwidth, clip=true, trim={2cm 0cm 1cm 0cm}]{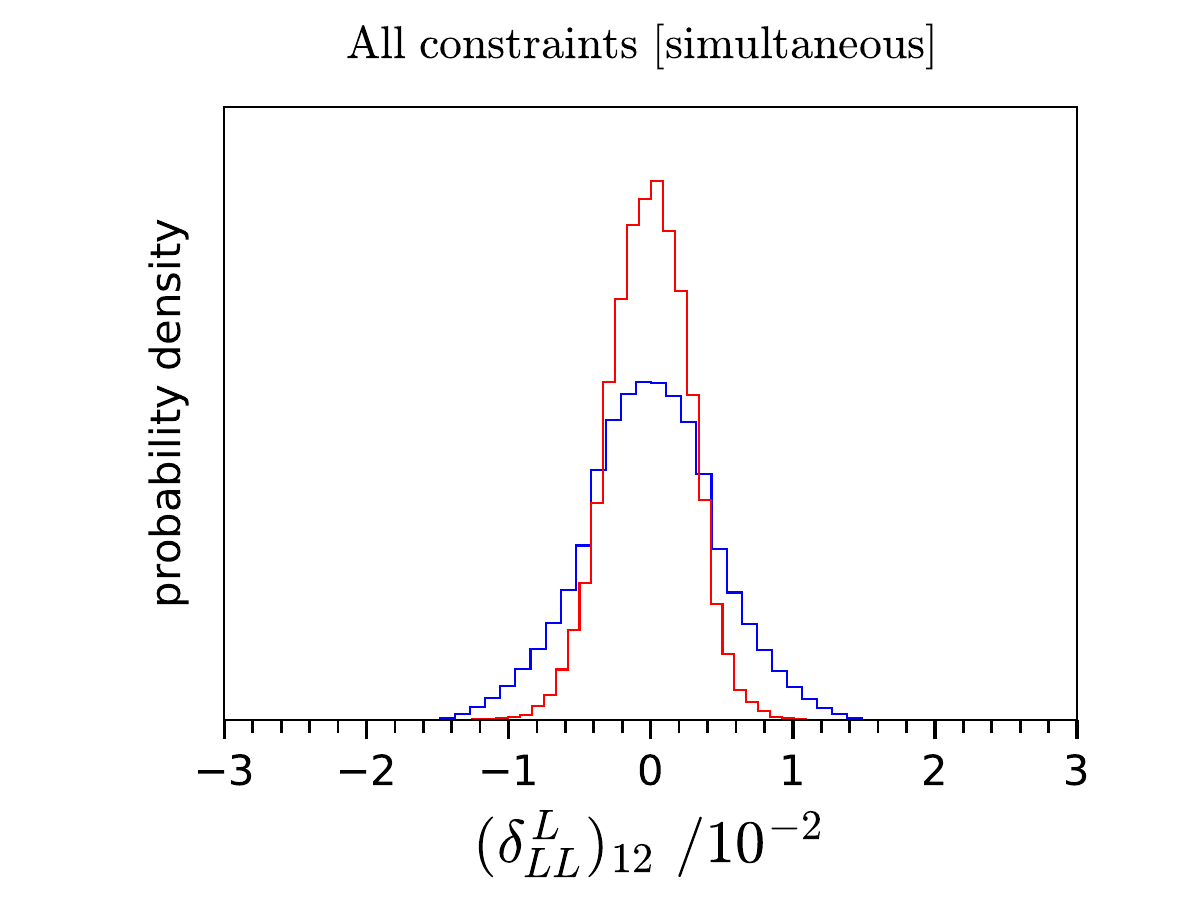}
		\includegraphics[width=0.495\textwidth, clip=true, trim={2cm 0cm 1cm 0cm}]{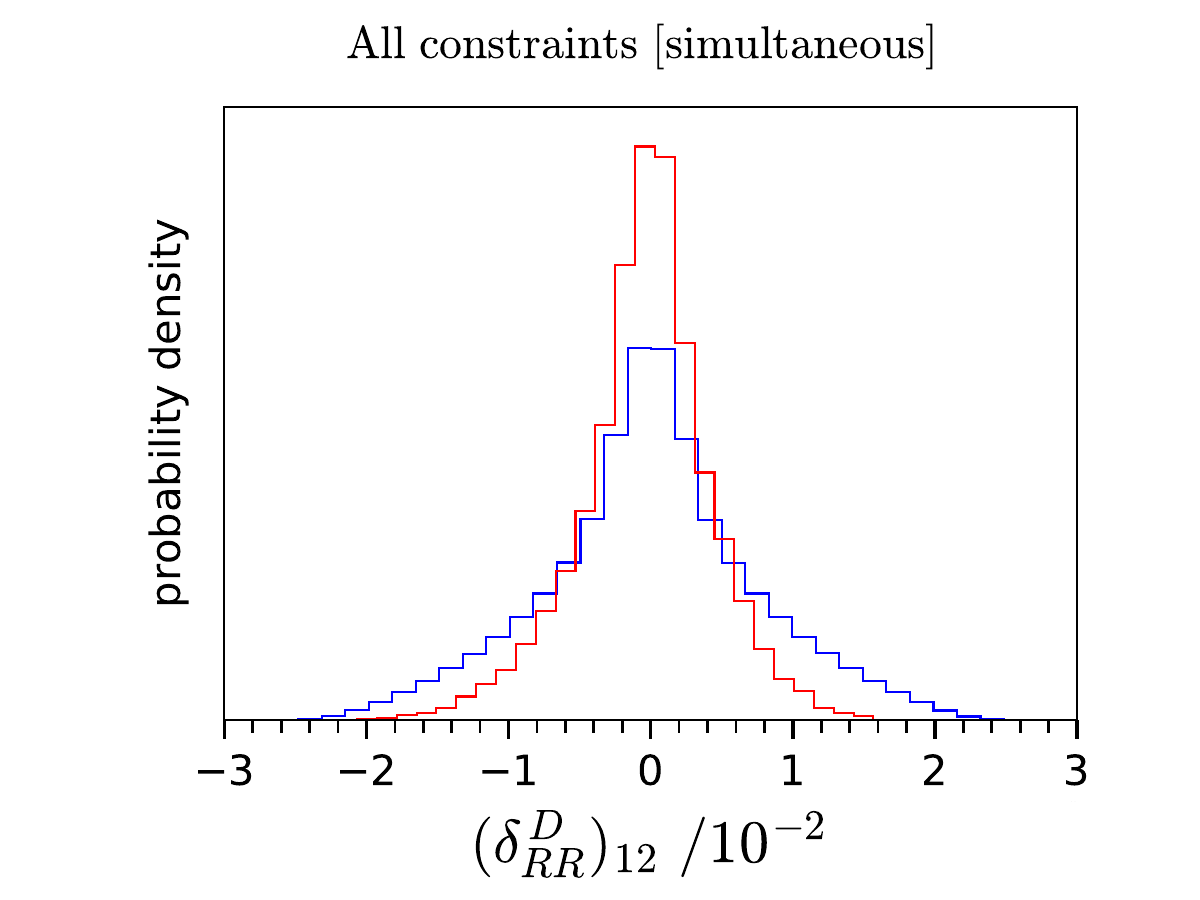}};
		\node[anchor=south west,inner sep=0] at (0,-13.6) {	
		\includegraphics[width=0.495\textwidth, clip=true, trim={2cm 0cm 1cm 0cm}]{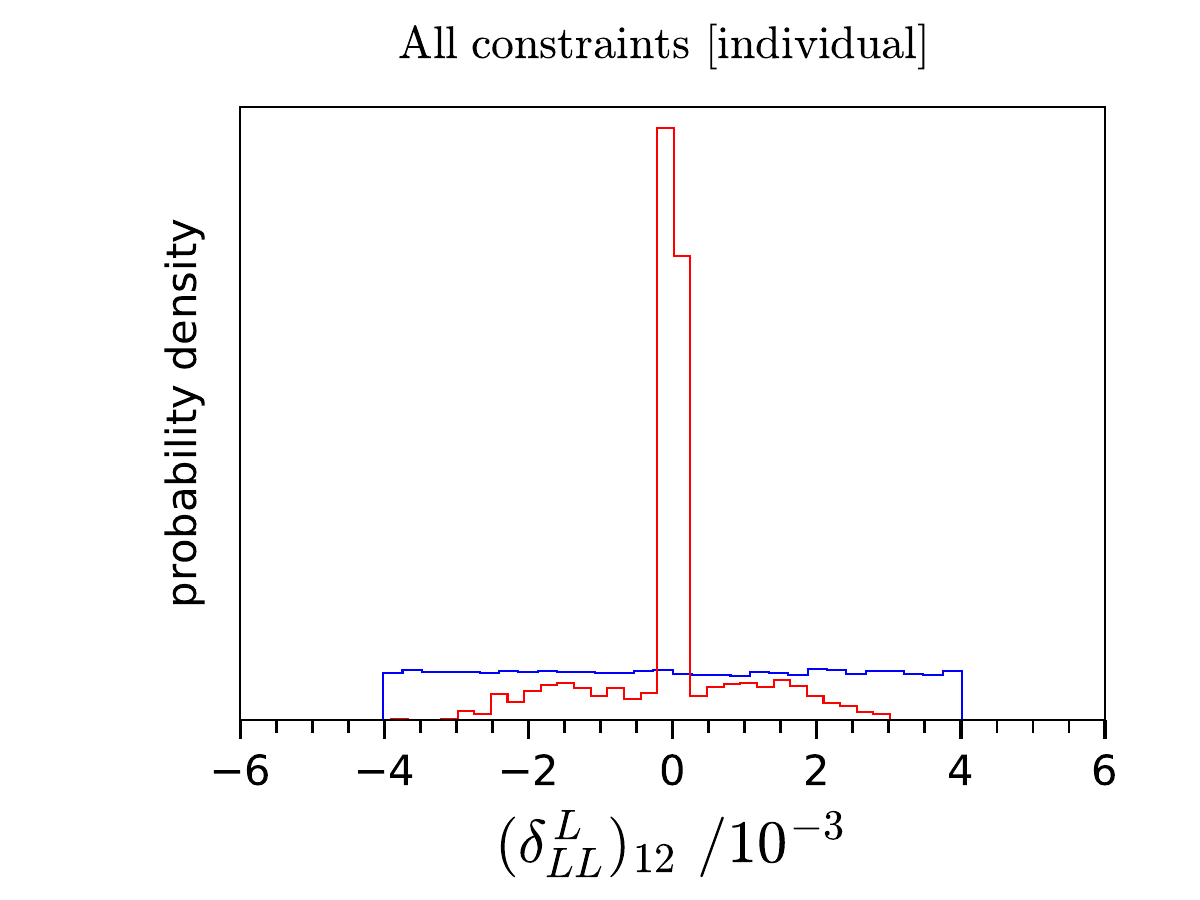}
		\includegraphics[width=0.495\textwidth, clip=true, trim={2cm 0cm 1cm 0cm}]{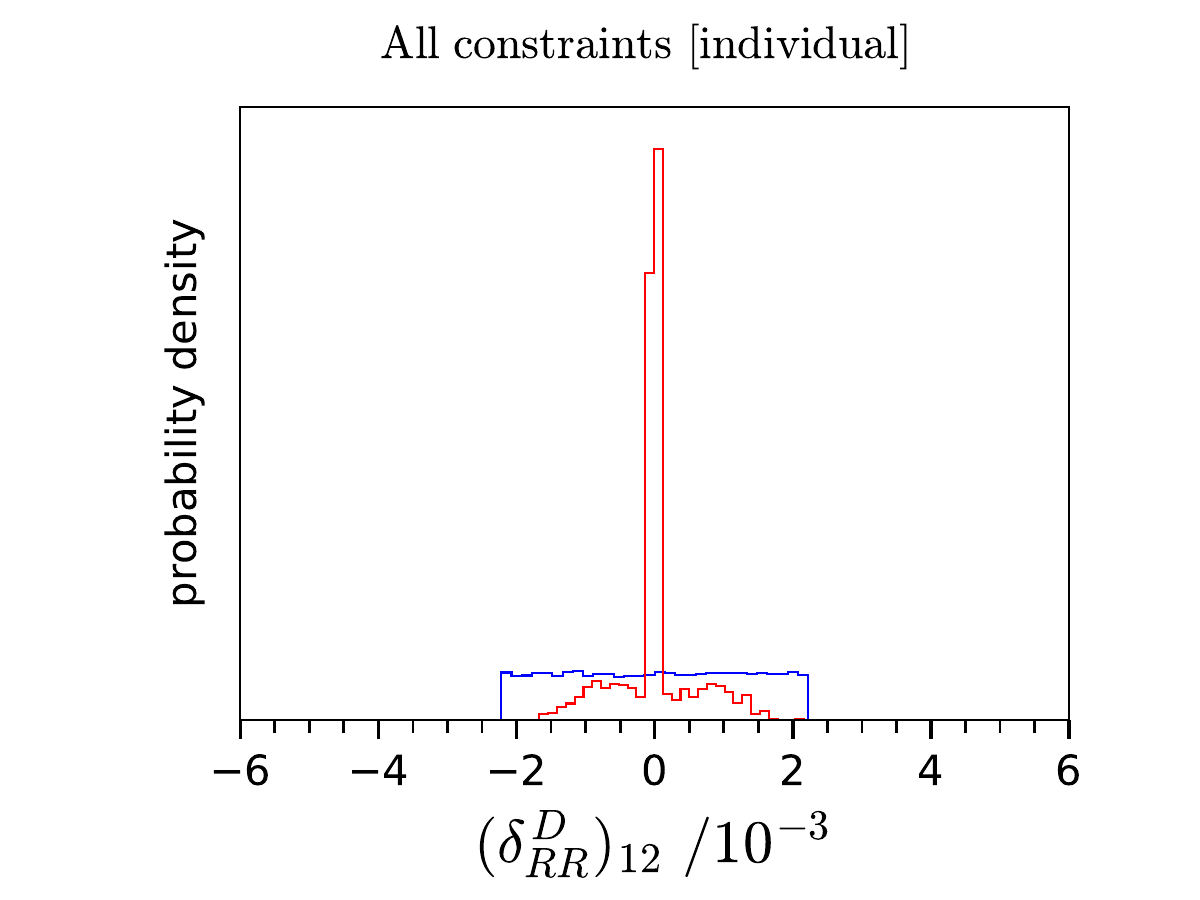}};
		\node at (4.7,5.5) {\textbf{a)}};
		\node at (1.1,-1.3) {\textbf{b)}};
		\node at (8.7,-1.3) {\textbf{c)}};
		\node at (1.2,-8.1) {\textbf{d)}};
		\node at (8.8,-8.1) {\textbf{e)}};
	\end{tikzpicture}
	\caption{Distributions obtained for the GUT-scale parameter $(\deltaf)_{12}$ and the associated SUSY-scale parameters $(\delta^{L}_{LL})_{12}$ and $(\delta^{D}_{RR})_{12}$ (see Eq.\ \eqref{Eq:MSSM_small_deltas}) from simultaneous (b) and c)) and individual (d) and e)) scan around Scenario 2. Analogously to other results, prior distributions are shown in blue and posterior distributions are shown in red.}
	\label{fig:GUT_to_SUSY_delta_parameters}
\end{figure}
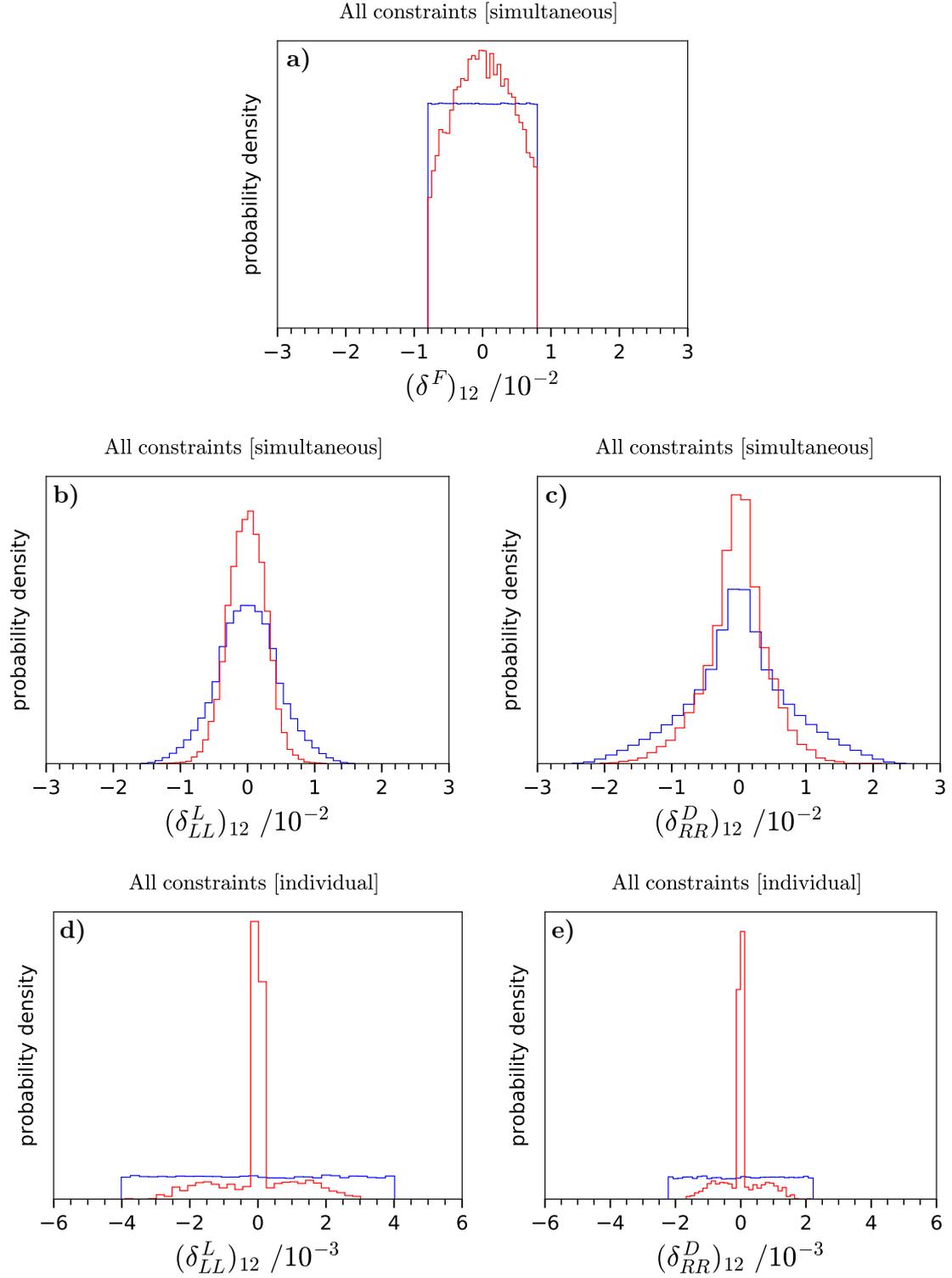

In Fig.\ \ref{fig:GUT_to_SUSY_delta_parameters} we show the example of $(\deltaf)_{12}$, defined at the GUT scale, and the two resulting SUSY scale parameters $(\delta^L_{LL})_{12}$ and $(\delta^D_{RR})_{12}$, which belong to the slepton and down-type squark sectors, respectively. First, we see that the prior distribution is altered by the renormalization group effects between the GUT scale in panel a) and the SUSY scale distributions in panels b) and c). The imposed flat priors at the GUT scale are transformed into almost Gaussian-like distributions at the SUSY scale. Looking at the corresponding posteriors, the SUSY scale distributions look even more peaked than the corresponding GUT scale histrograms. 

Second, it is interesting to note that, at the SUSY scale, the allowed range for the hadronic parameter $(\delta^{D}_{RR})_{12}$ is wider than that for the related leptonic parameter $(\delta^{L}_{LL})_{12}$ in the simultaneous scan. This behaviour is somewhat unexpected, since the gluino running, which is blind to flavour, drives the diagonal squark mass parameters higher, while it leaves the leptonic ones unaffected. In turn, this is expected to reduce the squark NMFV parameters once normalized as per Eq.\ \eqref{Eq:MSSM_small_deltas} \cite{Ciuchini:2007ha}. We find that this behaviour is confirmed for all NMFV parameters stemming from {\it individual} scans (see examples in Fig.\ \ref{fig:GUT_to_SUSY_delta_parameters} panels d) and e)), agreeing with the results presented in Ref.\ \cite{Ciuchini:2007ha}. However, for the $\deltaf$ parameters, the reverse is true when considering the {\it simultaneous} scan. We suspect that strong renormalization group effects are the cause of this feature, due to the fact that multiple NMFV parameters interact with each other during the evolution from the GUT scale to the SUSY scale.

\subsection{Parameter Correlations}
\label{sec:param_correlations}

In this section, we examine more closely the correlation between certain NMFV parameters, mentioned already several times in the above discussion, and being the reason that scanning over all parameters simultaneously is ultimately required. The key is that cancellations may exist between the contributions from certain parameters in the calculation of a given observable. However, dealing with analytical results for the different experimental constraints is difficult and beyond the scope of this work. Instead, we choose to take advantage of the numerical results, showing posterior distributions of more than one NMFV parameter together. 

\begin{figure}
	\centering
	\includegraphics[width=0.495\textwidth, clip=true, trim={1cm 0cm 1cm 0cm}]{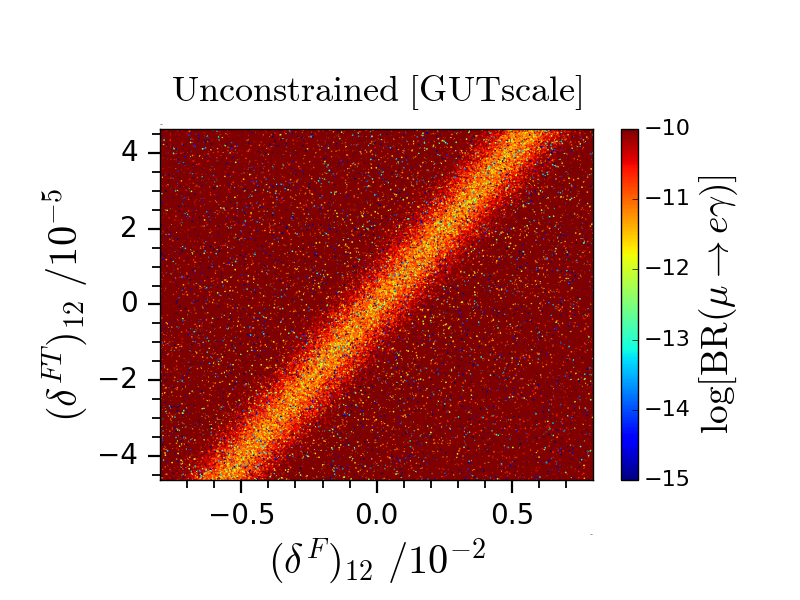}
	\includegraphics[width=0.495\textwidth, clip=true, trim={0.5cm 0cm 1cm 0cm}]{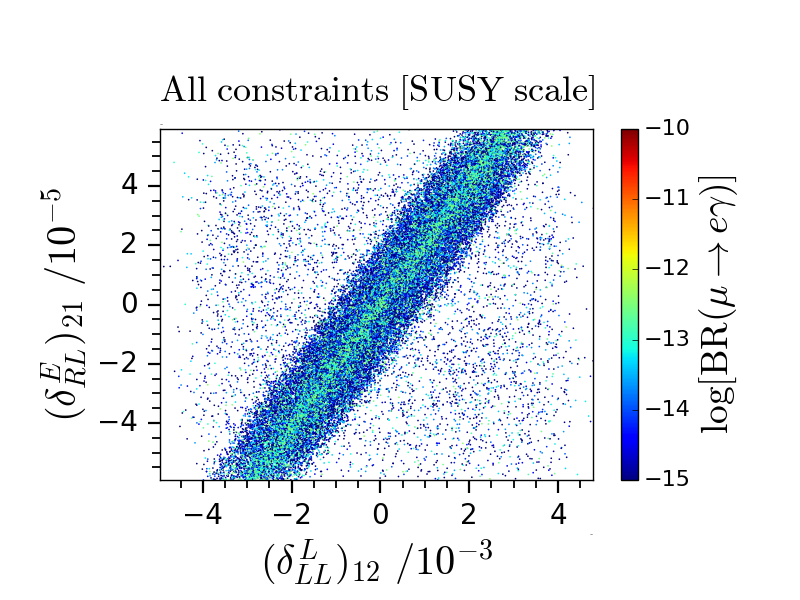}
	\caption{Correlation of the GUT-scale parameters $(\delta^F)_{12}$ and $(\delta^{FT})_{12}$ (left panel) and associated correlation of the SUSY-scale parameters $(\delta^L_{LL})_{12}$ and $(\delta^E_{RL})_{12}$ (right panel) for Scenario 1. While the first plot shows the results for the full scan, the second one shows only the surviving points once the constraints of Table \ref{Tab:Constraints} are applied.}
	\label{Fig:CorrelationsPlots1}
\end{figure}

\begin{figure}
	\centering
	\includegraphics[width=0.495\textwidth, clip=true, trim={1cm 0cm 1cm 0cm}]{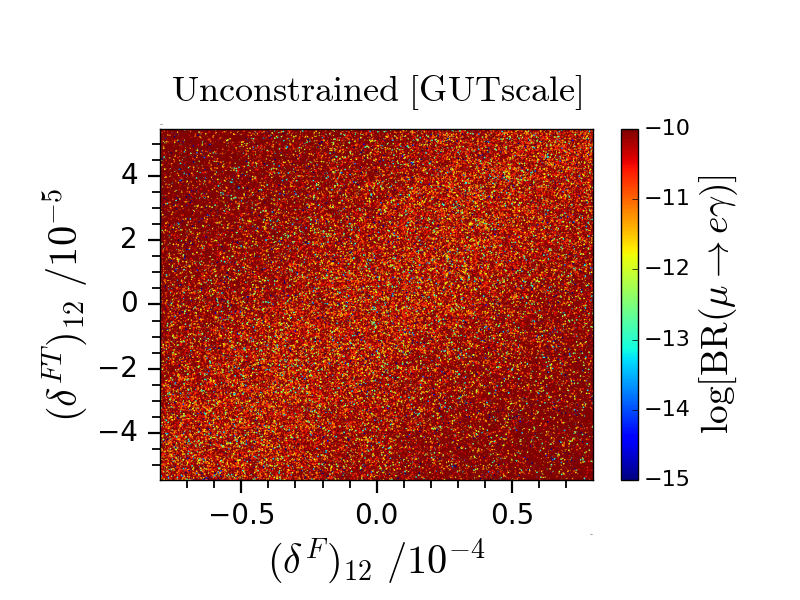}
	\includegraphics[width=0.495\textwidth, clip=true, trim={0.5cm 0cm 1cm 0cm}]{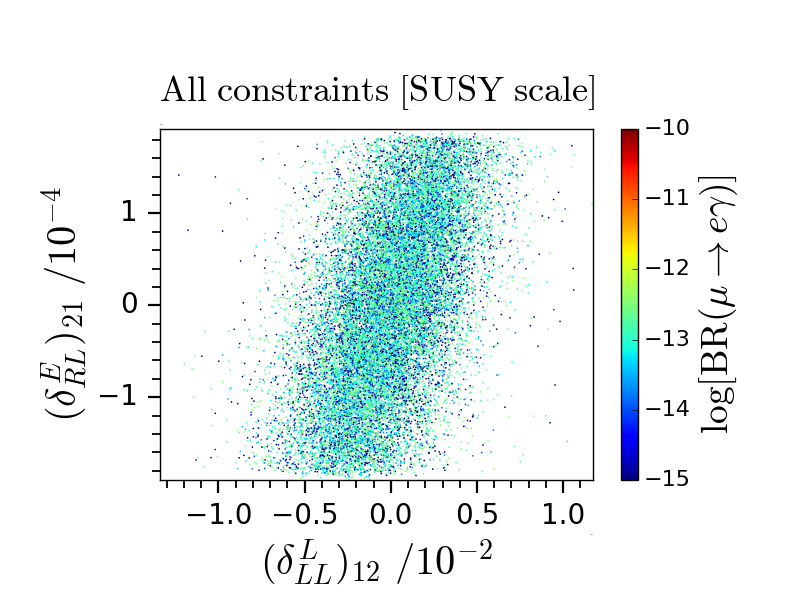}
	\caption{Correlation of the GUT-scale parameters $(\delta^F)_{12}$ and $(\delta^{FT})_{12}$ (left panel) and associated correlation of the SUSY-scale parameters $(\delta^L_{LL})_{12}$ and $(\delta^E_{RL})_{12}$ (right panel) for Scenario 2. While the first plot shows the results for the full scan, the second one shows only the surviving points once the constraints of Table \ref{Tab:Constraints} are applied.}
	\label{Fig:CorrelationsPlots2}
\end{figure}

The first panel in Fig.\ \ref{Fig:CorrelationsPlots1} shows viable parameter points that seem to follow a ``golden line'', with an increased density of points concentrated around a linear relationship between the GUT scale parameters $(\deltaf)_{12}$ and $(\deltaft)_{12}$. Indeed, the impact of $\mathrm{BR}(\mu \rightarrow e \gamma)$ is suppressed in this line due to cancellation between the two parameters in the analytic expression for this observable. One can also see this in the right panel, only those points lying close to or along said correlation line are consistent with the experimental limits. Said correlation could provide an interesting hint for future SUSY GUT model building.

The analytic expression for the decay rate of $\mu \rightarrow e \gamma$ can be written as \cite{Ciuchini:2007ha}
\begin{align}
	\frac{\mathrm{BR}(\ell_i\rightarrow \ell_j \gamma)}{\mathrm{BR}(\ell_i\rightarrow \ell_j \nu_i \nu_j)} ~=~
	\frac{48\pi^3\alpha}{G_{F}^2}\big( |F_{L}^{ij}|^2 + |F_{R}^{ij}|^2  \big)
	\label{eqn:analytic_muegamma}
\end{align}
where the branching ratio of the decay $\ell_i\rightarrow \ell_j \nu_i \nu_j$ is a constant with respect to the NMFV parameters under consideration in the present work. For real NMFV parameters, the form factors $F_{L,R}$ are related to the flavour violating parameters at the SUSY scale according to
\begin{align}
	\begin{split}
		F_{L}^{ij} ~&=~ c_1 (\delta_{LL}^L)_{ij} + c_2 (\delta_{RL}^E)_{ij} \,, \\
		F_{R}^{ij} ~&=~ c_3 (\delta_{RR}^L)_{ij} + c_4 (\delta_{RL}^E)_{ji} \,.
	\end{split}
	\label{eqn:a_params}
\end{align}
The coefficients $c_i$ ($i=1,\ldots,4$) are combinations of loop factors, masses, and other numerical inputs which can be assumed to be constant in our analysis. Minimizing the form factors $F_{L,R}$ in Eq.\ \eqref{eqn:a_params} to yield small $\mu\rightarrow e\gamma$ branching ratios and hence satisfy the experimental constraint leads to relations of the form
\begin{align}
	(\delta_{LL}^L)_{ij} ~=~  -\frac{2c_2}{c_1} \, (\delta_{RL}^E)_{ij} \,,
	\label{eqn:correlation_line_equation}
\end{align}
corresponding to the observed lines in Figs.\ \ref{Fig:CorrelationsPlots1} and \ref{Fig:CorrelationsPlots2}. As such, the ``golden line'' that we recover purely from our numerical analysis is consistent with the analytic formulae for this lepton flavour-violating decay.

\section{Conclusion}
\label{Sec:Conclusion}

In this paper we have considered CP-conserving 
non-minimal flavour violation in $A_4\times SU(5)$ inspired Supersymmetric Grand Unified Theories (GUTs),
focussing on the regions of parameter space where Dark Matter is successfully accommodated due to a light right-handed
smuon a few GeV heavier than the lightest neutralino dark matter candidate. Such regions of parameter space 
are obtained by choosing the second generation $T_2$ to have a light soft mass, while the heavy gluino mass
ensured that all squarks in this multiplet are heavy after RG running to low energy.
We have considered two scenarios along those lines, one with a very light right-handed smuon, which is capable of being discovered or excluded by the LHC very soon, but which can account for the 
$(g-2)_{\mu}$ results, and another scenario with a somewhat heavier smuon.
In such regions of parameter space we have found that some of the flavour violating parameters, in particular $(\delta^T)_{13}$ and $(\delta^{FT})_{32}$, are constrained by the requirement of 
dark matter relic density, due to the delicate interplay between the smuon and neutralino masses.

By scanning over many of the GUT scale flavour violating parameters, constrained by low energy quark and lepton flavour violating observables, we have discovered a striking 
difference between the results in which individual parameters are varied to those where multiple parameters are 
varied simultaneously, where the latter relaxes the constraints on flavour violating parameters due to 
cancellations and/or correlations. 
Since charged lepton flavour violation provides the strongest constraints within a GUT framework, 
due to relations between quark and lepton flavour violation, we have examined in detail 
a prominent correlation between the flavour violating parameters $(\delta^{F})_{12}$ 
and $(\delta^{FT})_{12}$
at the GUT scale consistent with the
stringent lepton flavour violating process $\mu \rightarrow e \gamma$.
By switching on both flavour violating parameters together, we have seen that much larger flavour violation
is allowed than if only one of them were permitted separately. 
We have examined this correlation also in terms of the resulting low energy flavour violating parameters in the quark and 
lepton sectors, and have provided some analytic estimates to understand the origin of the observed correlation.

Precision flavour physics measurements could present challenges to this work and warrant further attention. Particularly, situations such as this often predict small-but-non-zero branching ratios for the LFV decays $\mu\rightarrow e\gamma$ and $\mu\rightarrow 3e$, hence stricter bounds on such processes will further limit the amount of NMFV allowed in such scenarios. Figs.\ \ref{Fig:CorrelationsPlots1} and \ref{Fig:CorrelationsPlots2} are purely data-driven and shows the regions that experimental data prefers; a model which predicts such a correlation could allow reasonable flavour violation and still be preferred over other such models.

In general, we have examined
the relation between GUT scale and low scale flavour violating parameters, for both quarks and leptons, 
and shown how the usual expectations may be violated due to the correlations when multiple parameters are 
varied simultaneously. We have presented results in the 
framework of non-minimal flavour violation in  $A_4\times SU(5)$ inspired Supersymmetric Grand Unified Theories, with smuon assisted dark matter. Such a framework is interesting since 
it allows both successful dark matter and contributions to $(g-2)_{\mu}$, as well as providing the smoking gun prediction of a light right-handed smuon accessible at LHC energies.

\acknowledgments
S.\,F.\,K.\ acknowledges the STFC Consolidated Grant ST/L000296/1 and the European Union's Horizon 2020 Research and Innovation programme under Marie Sk\l{}odowska-Curie grant agreements Elusives ITN No.\ 674896 and InvisiblesPlus RISE No.\ 690575.
The authors acknowledge the use of the IRIDIS High Performance Computing Facility, and associated support services at the University of Southampton, in the completion of this work.
The work of J.\,B.\ is funded by a PhD fellowship by the French Ministery for Education and Research. J.\,B. acknowledges the Universite Savoie Mont Blanc for the project specific funding provided by their {\it Aide \`a la mobilit\'e internationale} program. The work of J.\,B.\ and B.\,H.\ is supported by {\it Investissements d'avenir}, Labex ENIGMASS, contrat ANR-11-LABX-0012. 
S.\,J.\,R.\, is supported by a Mayflower studentship from the University of Southampton.
The figures presented in this paper have been generated using {\tt MatPlotLib} \cite{Matplotlib}.


\appendix 

\section{SPheno and the SCKM basis}
\label{Appendix:SCKM}

The CKM basis is the one in which the up- and down-type quark Yukawa matrices are diagonal. The Super-CKM basis (SCKM) is obtained analogously, i.e.\ the squarks undergo the same rotations as their SM partners. This basis is convenient for phenomenological studies, and allows for a consistent expression of flavour vioation throughout the literature. The different rotations for the SM quark and lepton fields are:
\begin{align}
	\begin{split}
		u_{L}' = V_{u_L} u_{L} \,,\quad u_{R}' &= V_{u_R} u_R \,,\quad d_{L}' = V_{d_L} d_L \,,\quad d_{R}' = V_{d_R} d_R \,, \\ 
		e_{L}' &= V_{e_L} e_L \,,\quad~ e_{R}' = V_{e_R} e_R \,,
	\end{split}
\end{align}
where the primed fields are in the flavour basis and the bare fields are in the basis of diagonal Yukawa couplings. The misalignment between up- and down-type quarks leads to the usual CKM matrix:
\begin{align}
	V_{\mathrm{CKM}}=V_{u_L}^\dagger V_{d_L}
\end{align}
In order to account for the change to the SCKM basis, the numerical programme {\tt SPheno} assumes diagonal down-type Yukawa matrices. In this case the CKM matrix is the following: $V_{\mathrm{CKM}}=V_{u_L}^\dagger\mathcal{I}=V_{u_L}^\dagger$.

In $SU(5)$-like models, the choice of the representations $F=\bf{\bar{5}}$ and $T={\bf 10}$ forces relationships between Yukawa couplings to hold at the unification scale:
\begin{align}
	y_u = y_{u}^T \ \ \ \mathrm{and} \ \ \ y_d=y_{e}^T.
\end{align}
As a consequence, we have $V_{u_L} = V_{u_R}$ in this case, meaning both lepton and down-type Yukawas are simultaneously diagonal. For consistency, we then have to perform a systematic CKM rotation for all terms involving $V_{u_L}$ and $V_{u_R}$. The soft-breaking terms of the Lagrangian transform as follows when switching to the SCKM basis:
\begin{align}
	\overline{\widetilde{U}'_{L,R}} \, M_{T}^2 \, \widetilde{U}_{L,R}' ~&=~ 
		\overline{\widetilde{U}_{L,R}} \, V_{\mathrm{CKM}} \, M_{T}^2 \, V_{\mathrm{CKM}}^\dagger \, \widetilde{U}_{L,R} \,,  
		\nonumber \\
	\overline{\widetilde{U}'_{R}} \, A_u \widetilde{U}_{L}' ~&=~ 
		\overline{\widetilde{U}_{R}} \, V_{\mathrm{CKM}} \, A_u \, V_{\mathrm{CKM}}^\dagger \, \widetilde{U}_{L} \,, 
		\nonumber \\
	\overline{\widetilde{D}'_{L,R}} \, M_{T,F}^2 \, \widetilde{D}'_{L,R} ~&=~ 
		\overline{\widetilde{D}_{L,R}} \, M_{T,F}^2  \, \widetilde{D}_{L,R} \,,  
		\nonumber \\
	\overline{\widetilde{D}'_{R}} \, A_d \, \widetilde{D}'_{L} ~&=~ 
		\overline{\widetilde{D}_{R}} \, A_d \, \widetilde{D}_{L} \,, 
		\\
	\overline{\widetilde{L}'_{L}} \, M_{F,T}^2 \, \widetilde{L}'_{L} ~&=~ 
		\overline{\widetilde{L}_{L}} \, M_{F,T}^2 \, \widetilde{L}_{L} \,, 
		\nonumber \\
	\overline{\widetilde{E}'_{R}} \, M_{F,T}^2 \, \widetilde{E}'_{R} ~&=~ 
		\overline{\widetilde{E}_{R}} \, M_{F,T}^2 \, \widetilde{E}_{R} \,, 
		\nonumber \\
	\overline{\widetilde{E}'_{R}} \, A_{d}^T \, \widetilde{L}'_{L} ~&=~ 
		\overline{\widetilde{E}_{R}} \, A_{d}^T \, \widetilde{L}_{L} \,. 
		\nonumber
\end{align}
Consequently, in the SCKM basis, where the down-type Yukawa matrix is diagonal following the {\tt SPheno} requirements and assuming the $SU(5)$ relations, the $6\times6$ soft mass matrices in the MSSM (once trilinear couplings have been taken into account) are
\begin{gather}
	M_{\widetilde{D}}^2 = \begin{pmatrix}
		M_{T}^2 & \frac{v_d}{\sqrt{2}}A_{D}^T \\
		\frac{v_d}{\sqrt{2}}A_D & M_{F}^2
	\end{pmatrix} \,,
	\qquad\qquad
	M_{\widetilde{L}}^2 = \begin{pmatrix}
		M_{F}^2 & \frac{v_d}{\sqrt{2}}A_{D}\\
		\frac{v_d}{\sqrt{2}}A_D^T  & M_{T}^2
	\end{pmatrix} \,, \\
	M_{\widetilde{U}}^2 = \begin{pmatrix}
		V_{\mathrm{CKM}} M_{T}^2 V_{\mathrm{CKM}}^\dagger & \frac{v_u}{\sqrt{2}}V_{\mathrm{CKM}} A_{U}^T V_{\mathrm{CKM}}^\dagger \\ 
		\frac{v_u}{\sqrt{2}}V_{\mathrm{CKM}} A_{U} V_{\mathrm{CKM}}^\dagger & V_{\mathrm{CKM}} M_{T}^2 V_{\mathrm{CKM}}^\dagger
	\end{pmatrix} \,, \nonumber
\end{gather}
up to the $D$-terms and SM masses. Since {\tt SPheno} automatically ensures the CKM rotation for the left-left block of $M_{U}^2$, we enforce the rotation for the other blocks of the up-type squark mass matrix by hand before running {\tt SPheno}.


\end{document}